\documentclass[acmtocl]{acmtrans2m}

 \usepackage{algorithmic}
\usepackage{amssymb,amsfonts,amsmath}
\usepackage{graphicx,epsfig,boxedminipage,url,xspace,array}
\usepackage{pstricks,pst-node,pst-tree}
\usepackage{changebar}

\newcommand{\true}{\text{\normalfont\small\sffamily true}\xspace}
\newcommand{\false}{\text{\normalfont\small\sffamily false}\xspace}

\newcommand{\lhint}[1]{\text{(#1)}}

\usepackage[latin1]{inputenc}
  {\noindent\textbf{Proof:}\newline}%
  {\hfill {\framebox(5, 5)[!t]{}}}
\providecommand{\LyX}{L\kern-.1667em\lower.25em\hbox{Y}\kern-.125emX\@}
\newcommand{\Chi}{\raisebox{2pt}{$\chi$}}    %

\newcommand{\A}[1]{\forall #1 \cdot}
\newcommand{\E}[1]{\exists #1 \cdot}

\newcommand{\latstyle}[1]{\ensuremath{\mathbf{#1}}}

\newcommand{\limp}{\Rightarrow}

\newcommand{\Land}{\bigwedge}
\newcommand{\Lor}{\bigvee}
\newcommand{\lpmi}{\Leftarrow}
\newcommand{\liff}{\Leftrightarrow}

\newcommand{\B}{\latstyle{2}}
\newcommand{\Bisim}{\mathcal{B}}

\newcommand{\subst}{\leftarrow}

\newcommand{\kripke}{Kripke\xspace}

\newcommand{\ctl}{\ensuremath{\mathrm{CTL}}\xspace}
\newcommand{\qctl}{\ensuremath{\mathrm{QCTL}}\xspace}

\newcommand{\ctlast}{\ensuremath{\mathrm{CTL}^\ast}\xspace}
\newcommand{\qctlast}{\ensuremath{\mathrm{QCTL}^\ast}\xspace}
\newcommand{\actlast}{\ensuremath{\mathrm{ACTL}^\ast}\xspace}
\newcommand{\ectlast}{\ensuremath{\mathrm{ECTL}^\ast}\xspace}

\newcommand{\ltl}{\mbox{LTL}\xspace}

\makeatletter
\newcommand{\@jux}[1]{#1 \mskip-5.5mu #1}
\newcommand{\@jux@}[1]{#1 \mskip-2.6mu #1}
\newcommand{\mv}[1]{\@jux@{[} #1 \@jux@{]}}
\newcommand{\mvB}[1]{\@jux@{[} #1 \@jux@{]}^B}
\@ifundefined{bigsqcap}
  {}
  {\relax}

\def\pspicgrid(#1,#2)(#3,#4){%
\begingroup\pst@@@picture[](#1,#2)(#3,#4)%
\psgrid[subgriddiv=1,griddots=10,gridlabels=7pt](#1,#2)(#3,#4)}

\makeatother

\newcommand{\eqdef}{\triangleq}

\newcolumntype{H}{>{$\hspace{0.1in}\normalfont\small(}l<{)$}}

\newcommand{\AG}{\mathopen{AG}}
\newcommand{\EG}{\mathopen{EG}}

\newcommand{\AF}{\mathopen{AF}}
\newcommand{\EF}{\mathopen{EF}}

\newcommand{\AX}{\mathopen{AX}}
\newcommand{\EX}{\mathopen{EX}}
\newcommand{\U}{\mathbin{U}}
\newcommand{\dU}{\mathbin{\tilde{U}}}

\newcommand{\mL}{\mathcal{L}}
\newcommand{\mM}{\mathcal{M}}
\newcommand{\mN}{\mathcal{N}}
\newcommand{\mO}{\mathcal{O}}
\newcommand{\mP}{\mathcal{P}}
\newcommand{\mQ}{\mathcal{Q}}
\newcommand{\mU}{\mathcal{U}}
\newcommand{\mV}{\mathcal{V}}
\newcommand{\mX}{\mathcal{X}}
\newcommand{\ez}{\textit{ez}}
\newcommand{\variant}[2]{{#1}|_{#2}}

\newcommand{\tuple}[1]{\langle #1 \rangle}
\newcommand{\setof}[1]{\{ #1 \}}

\newcommand{\heading}[1]{\vspace{0.2in}\noindent\textbf{#1}\hspace{0.05in}}

\newtheorem{theorem}{Theorem}[section]

\newtheorem{corollary}[theorem]{Corollary}
\newtheorem{proposition}[theorem]{Proposition}

\newdef{definition}[theorem]{Definition}
\newdef{remark}[theorem]{Remark}

\title{Robust Vacuity for Branching Temporal Logic}
\author{ARIE GURFINKEL \\Software Engineering Institute, Carnegie Mellon University \and MARSHA CHECHIK \\ University of Toronto}
\begin{abstract}
  There is a growing interest in techniques for detecting whether a
logic specification is satisfied too easily, or \emph{vacuously}.  For
example, the specification ``every request is eventually followed by
an acknowledgment'' is satisfied vacuously by a system that never
generates any requests. Vacuous satisfaction misleads users of
model-checking into thinking that a system is correct. It is a serious
problem in practice.

There are several existing definitions of vacuity. Originally, Beer et
al.  formalized \emph{vacuity} as insensitivity to syntactic
perturbation (\emph{syntactic vacuity}). This formulation captures the
intuition of ``vacuity'' when applied to a single occurrence of a
subformula. Armoni et al. argued that vacuity must be \emph{robust} --
not affected by semantically invariant changes, such as extending a
model with additional atomic propositions. They show that syntactic
vacuity is not robust for subformulas of linear temporal logic, and
propose an alternative definition -- \emph{trace vacuity}.

In this article, we continue this line of research. We show that trace
vacuity is not robust for branching time logic. We further refine the
notion of vacuity so that it applies uniformly to linear and branching
time logic and does not suffer from the common pitfalls of prior
definitions. Our new definition -- \emph{bisimulation vacuity} -- is a
proper and non-trivial extension of both syntactic and trace
vacuity. We discuss the complexity of detecting bisimulation vacuity,
and identify several practically-relevant subsets of \ctlast for which
vacuity detection problem is reducible to model-checking. We believe
that in most practical applications, bisimulation vacuity provides
both the desired theoretical properties and is tractable
computationally.


\end{abstract}

\def\firstfoot{\def\@firstfoot{}}
\def\runningfoot{\def\@runningfoot{}}

\begin{document}
\maketitle

\section{Introduction}
\label{sec:introduction}

Model-checking gained wide popularity as an automated technique for
effective analysis of software and hardware systems.  Given a temporal
logic property, the model-checker automatically determines whether the property is satisfied by the system, giving a counterexample in case of the failure.

Yet a major
problem in practical applications of model-checking is that a
successful run of the model-checker does not necessarily guarantee
that the intended requirement is satisfied by the system~\cite{beer97,beatty94}.
For example, consider the property
\begin{quote}``every request must be followed by an acknowledgment'',
\end{quote}
where the environment controls the requests.  This property, expressed
in \ctl as $\AG (\text{req} \limp \AF \text{ack})$, is satisfied
vacuously\footnote{Beatty and Briant~\cite{beatty94} originally called
  this problem ``antecedent failure''.} by any system that never
produces a request (i.e., $\text{req}$ is false in all reachable
states). In this case, the environment alone ensures satisfaction of
this property, so it is true of \emph{any} system combined with such
an environment.  Intuitively, a property $\varphi$ is considered
vacuous if it contains a subformula that is irrelevant for $\varphi$'s
satisfaction by the system.  In the above example, it is $\AF
\text{ack}$.

Researchers at the IBM Haifa Research Laboratory observed that vacuity
is a serious problem~\cite{beer97} and that ``... typically 20\% of
specifications pass vacuously during the first formal verification
runs of a new hardware design, and that vacuous passes always point to
a real problem in either the design, or its specification, or the
environment''~\cite{beer97}.  Further justification has been given by
several researchers, such as the case study by Purandare and
Somenzi~\cite{purandare02}.  These results led to a substantial
interest in techniques for detecting vacuity.

Most of the early work on vacuity detection
uses a \emph{syntactic} definition of vacuity, provided by Beer et
al.~\cite{beer01}: a formula $\varphi$ is syntactically vacuous in a
subformula $\psi$ and model $K$, if replacing $\psi$ by any other
temporal logic formula $x$, denoted $\varphi[\psi \subst x]$, does not
affect the satisfaction of $\varphi$ in $K$. That is, $\varphi$ is
vacuous if $\A{x \in TL} \varphi[\psi \subst x]$ is true, where $TL$
stands for a temporal logic. The main advantage of this definition is
the simplicity of detecting vacuity in an occurrence of a
subformula. That is, whenever $\psi$ occurs in $\varphi$ only once,
detecting whether $\varphi$ is syntactically vacuous in $\psi$ reduces
to model-checking $\varphi[\psi \subst \true]$ or $\varphi[\psi \subst
\false]$, based on the polarity of $\psi$.  This result started a line
of research, e.g.,
\cite{dong02,kupferman03,gurfinkel04a,bustan05,tzoerf06}, that aims to
increase the scope of applicability of vacuity detection algorithms.
In particular, this work deals with deciding vacuity for various
temporal logics, for formulas with one or multiple occurrences of a
subformula, handling vacuous satisfaction and vacuous failure of
formulas, and generating witnesses to non-vacuity.

An orthogonal question, raised by Armoni et al.~\cite{armoni03} and
continuing in this article, is to reexamine the \emph{meaning} of
vacuity.  Armoni et al. showed that the definition of syntactic
vacuity is too restrictive. It is not well suited for detecting
vacuity with respect to multiple occurrences of a subformula, i.e.,
deciding whether $(\AX p) \lor (\AX \neg p)$ is vacuous in
$p$. Furthermore, it is sensitive to irrelevant changes to the
model. For example, syntactic vacuity of a formula `if $p$ is true
now, it will remain true in the next state'', expressed in \ctl as
$\AG (p \limp \AX p)$), can be affected, i.e., changed from vacuous to
non-vacuous, by simply adding new atomic propositions to the model.

As an alternative, the authors of \cite{armoni03} develop a new definition, 
applicable to linear-time logic, called
\emph{trace vacuity}. Trace vacuity is not syntactic, but is based on
the semantics of quantified temporal logic. The new definition is
shown to alleviate the problems of syntactic vacuity (at least on the
examples tried by the authors). Furthermore, the complexity
of detecting vacuous satisfaction for LTL properties with respect to trace vacuity is in
the same complexity class as model-checking. 

In this article, we continue the search for the ``right'' definition of
vacuity, and whether this definition changes as we transition from
\ltl properties to \ctlast and from vacuous satisfaction (i.e.,
vacuity of formulas that are satisfied by the model) to vacuous
failure (i.e., vacuity of formulas that are violated by the model).
In particular, we develop a robust definition of vacuity, which we
call \emph{bisimulation} vacuity. We start with a definition of vacuity for
propositional logic, argue that it is robust, and then systematically
extend it to branching-time temporal logic \ctlast. We show that
bisimulation vacuity is a proper extension of syntactic vacuity: while
syntactic and bisimulation vacuity coincide for vacuity in a single
occurrence, syntactic vacuity is not robust when applied to vacuity in
multiple occurrences. Bisimulation vacuity is also a proper non-trivial
extension of trace vacuity: while the bisimulation and the trace vacuity
definitions coincide for \ltl, trace vacuity is not robust when
applied to branching-time logics.

We study the complexity of detecting bisimulation vacuity. In general,
this problem is EXPTIME-complete for \ctl and 2EXPTIME-complete for
\ctlast. However, we identify several important fragments of \ctlast
for which vacuity detection, or at least detecting vacuous
satisfaction, is no harder than model-checking.  In particular, we
show that checking vacuous satisfaction of A\ctlast is reducible to
model-checking, which subsumes the results of \cite{armoni03}.

The rest of the article is organized as follows.  We provide the
necessary background in Section~\ref{sec:background}. In
Section~\ref{sec:vacuity-definition}, we examine the meaning of
``robustness'' of vacuity, define bisimulation vacuity, and argue that it
is robust. In Section~\ref{sec:compl-vacu-detect}, we study complexity
of detecting bisimulation vacuity for \ctlast and identify subsets of
this language where this problem is tractable.  
We analyze the relationship between vacuity and abstraction
in Section~\ref{sec:vacuity-abstraction}.  We then compare our
approach with related work in Section~\ref{sec:related-work} and
conclude in Section~\ref{sec:conclusion}.

\vspace{-0.12in}

\section{Background}
\label{sec:background}
In this section, we give a brief overview of temporal logic
model-checking, property reserving relations, and several semantics of
quantified temporal logic.

\subsection{Models of Computation}
\label{sec:models}
  We use \kripke structures to model computations.  Intuitively, these
  are transition systems whose states are labeled by atomic
  propositions.  In this section, we review the formal definition of
  \kripke structures, and fix the notation.

We use $\latstyle{2}$ to
denote the set of boolean values $\{\true, \false\}$.
\begin{definition}[\kripke Structure]
  \label{def:kripke}
  A \emph{\kripke structure} $K$ is a tuple $(AP, S, R, s_0, I)$,
  where $AP$ is a set of atomic propositions, $S$ is a finite set of
  states, $R \subseteq S \times S$ is a total transition relation,
  $s_0 \in S$ is a designated initial state, and $I: S \to \B^{AP}$ is
  a labeling function, assigning a value to each atomic proposition $p
  \in AP$ in each state.
\end{definition}
Example \kripke structures are shown in
Figures~\ref{fig:k-one}~and~~\ref{fig:xvariants}. For two states $s$
and $t$, we write $R(s,t)$ for $(s,t) \in R$, and $R(s)$ to denote the
set of successors of $R$:
\begin{equation*}
  R(s) \eqdef \{t \in S \mid R(s, t)\}.
\end{equation*}
For notational convenience, we denote components of a \kripke
structure $K$ using the same typographical convention as used for $K$.
For example, $S'$ denotes the statespace of $K'$, $R'$ -- its
transition relation, $AP'$ -- the set of atomic propositions, etc.  A
\emph{path} $\pi$ of $K$ is an infinite sequence of states in which
every consecutive pair of states is related by the transition
relation. Let $i$ be a non-negative integer. We write $\pi(i)$ to
denote the $i+1$th state on the path, $\pi(0)$ to denote the first
state, and $\pi_i$ to denote the suffix of $\pi$ starting from the
$i$th state. The set of all paths of $K$ starting from a state $s$ is
denoted by $\Pi^K_s$ ($K$ is often omitted when clear from the
context).

We now define parallel synchronous composition.
\begin{definition}[Parallel Synchronous Composition]
  \label{def:sync-composition}
  Let $K_1 = (AP_1, S_1, R_1, s_1^0, I_1)$, and $K_2 = (AP_2, S_2,
  R_2, s_2^0, I_2)$ be two \kripke structures with disjoint atomic
  propositions, i.e., $AP_1 \cap AP_2 = \emptyset$. A \emph{parallel
    synchronous composition} of $K_1$ and $K_2$, written $K_1 || K_2$,
  is a \kripke structure $(AP_1 \cup AP_2, S_1 \times S_2, R_{||},
  (s_1^0, s_2^0), I_{||})$, where
  \begin{align*}
    R_{||}( (s, t), (s', t')) &\liff R_1(s, s') \land R_2(t, t')\\
    I_{||} ((s, t)) & \eqdef I_1(s) \cup I_2(t)\,.
  \end{align*}
\end{definition}

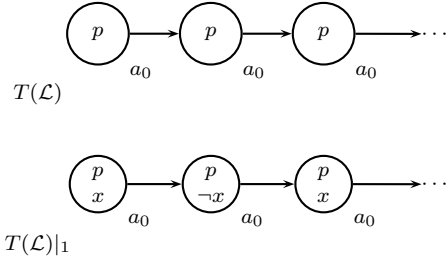
\begin{figure}[t]
  \centering
  \let\pspicgrid=\pspicture
\begin{pspicgrid}(0,0)(6,4)

\rput[bl](0,2){\begin{pspicture}(0,0)(6,2)
    \cnodeput[framesep=0.28](1,1){a}{$p$}
    \cnodeput[framesep=0.28](2.5,1){b}{$p$}
    \cnodeput[framesep=0.28](4,1){c}{$p$}
    \rput(5.5,1){\rnode{d}{$\cdots$}}
    
    \psset{arrows=->}
    \ncline{a}{b}
    \ncline{b}{c}
    \ncline{c}{d}

    \nput{-45}{a}{$a_0$}
    \nput{-45}{b}{$a_0$}
    \nput{-45}{c}{$a_0$}

    \rput(0.2,0.2){$T(\mL)$}

  \end{pspicture}}

\rput[bl](0,0){\begin{pspicture}(0,0)(6,2)
    \cnodeput[framesep=0](1,1){a}{\begin{tabular}{@{}c@{}}$p$\\$x$\end{tabular}}
    \cnodeput[framesep=-1pt](2.5,1){b}{\begin{tabular}{@{}c@{}}$p$\\$\neg x$\end{tabular}}
    \cnodeput[framesep=0](4,1){c}{\begin{tabular}{@{}c@{}}$p$\\$x$\end{tabular}}
    \rput(5.5,1){\rnode{d}{$\cdots$}}
    
    \psset{arrows=->}
    \ncline{a}{b}
    \ncline{b}{c}
    \ncline{c}{d}

    \nput{-45}{a}{$a_0$}
    \nput{-45}{b}{$a_0$}
    \nput{-45}{c}{$a_0$}

    \rput(0.2,0.2){$\variant{T(\mL)}{1}$}
  \end{pspicture}}

\end{pspicgrid}
  \caption{A tree unrolling $T(\mL)$ of $\mL$ 
    and one of its $\{x\}$-variant $\variant{T(\mL)}{1}$.}
  \label{fig:tree-unrolling}
\end{figure}

A computation tree $T(K)$ of a \kripke structure $K$ is an $S$-labeled
tree obtained by unrolling $K$ from its initial state. 
\begin{definition}[Computation Tree]
  \label{def:computation-tree}
  Let $K = (AP, S, R, s_0, I)$ be a \kripke structure. A
  \emph{computation tree} $T(K)$ of $K$ is an $S$-labeled tree $(T,
  \tau)$, where $T = (V, E)$ is a tree with vertex set $V$ and edge
  set $E$, and $\tau : V \to S$ is a labeling function, satisfying the
  ``unrolling'' conditions:
  \begin{enumerate}
   \item if $v$ is a root of $T(K)$, then $\tau (v) = s_0$;
    
   \item for a node $v$, $|E(v)| = |R(\tau(v))|$, and for each $s \in
    R(\tau(v))$ there exists a $u \in E(v)$ such that $\tau(u) = s$,
    where $E(v)$ is the set of successors of $v$.
  \end{enumerate}
\end{definition}
A tree unrolling $T(\mL)$ for a structure $\mL$ in
Figure~\ref{fig:k-one} is shown in Figure~\ref{fig:tree-unrolling}.
Note that since $\mL$ has only one transition, the unrolling is a
unary tree, i.e., a trace.

\begin{figure}[t]
  \centering
  \let\pspicgrid=\pspicture
\begin{pspicgrid}(0,0)(8,2)

\rput(1,1){\begin{pspicture}(0,0)(2,2)
  \cnodeput[framesep=0.28](1,1){a}{$p$}
  \psset{arrows=->}
  \nccircle[angleA=0]{->}{a}{.3}
  \nput*{-45}{a}{$a_0$}

  \rput(.5,0.3){$\mL$}
\end{pspicture}}

\rput(4,1){\begin{pspicture}(0,0)(2,2)
  \cnodeput[framesep=0pt](1,1){a}{\begin{tabular}{@{}c@{}}$p$\\$x$\end{tabular}}
  \psset{arrows=->}
  \nccircle[angleA=0]{->}{a}{.3}

  \rput(.5,0.3){$\variant{\mL}{1}$}
\end{pspicture}}

\rput(7,1){\begin{pspicture}(0,0)(2,2)
  \cnodeput[framesep=-1pt](1,1){a}{\begin{tabular}{@{}c@{}}$p$\\$\neg x$\end{tabular}}
  \psset{arrows=->}
  \nccircle[angleA=0]{->}{a}{.3}

  \rput(.5,0.3){$\variant{\mL}{2}$}
\end{pspicture}}

\end{pspicgrid}
  \caption{A \kripke structure $\mL$ and its $\{x\}$-variants $\variant{\mL}{1}$ and $\variant{\mL}{2}$.}
  \label{fig:k-one}
\end{figure}

\subsection{Temporal Logic}
\label{sec:tl}
Computation Tree Logic \ctlast~\cite{emerson85b} is a branching-time
temporal logic constructed from propositional connectives, temporal
operators $X$ (next), $U$ (until), $F$ (future), and $G$ (globally),
and path quantifiers $A$ (forall) and $E$ (exists).
\begin{definition}[Syntax of \ctlast]
  \label{def:ctlast-syntax}
  Temporal logic \ctlast denotes the set of all \emph{state formulas}
  satisfying the grammar
  \begin{equation*}
    \varphi \mathbin{::=} p \mid \varphi \land \varphi \mid
    \varphi \lor \varphi \mid \neg \varphi \mid A \psi \mid E \psi,
  \end{equation*}
  where $p$ is an atomic proposition, and $\psi$ is a \emph{path
    formula} satisfying the grammar
  \begin{equation*}
    \psi \mathbin{::=} \varphi \mid X \psi \mid \psi \U \psi \mid \psi \dU \psi \mid F \psi \mid G \psi.
  \end{equation*}
\end{definition}

The semantics of path formulas is given with respect to a path of a
\kripke structure. For a path formula $\psi$, we write $K, \pi \models
\psi$ to denote that $\psi$ is satisfied by the path $\pi$ of a
\kripke structure $K$. The semantics of state formulas is given with
respect to a state of a \kripke structure. For a state formula
$\varphi$, we write $K,s \models \varphi$ to denote that $\varphi$ is
satisfied in the state $s$ in $K$.
\begin{definition}[Semantics of \ctlast]
  \label{def:ctlast-semantics}
  Let $K = (AP, S, R, s_0, I)$ be a \kripke structure. The semantics
  of path and state formulas is defined as follows, where $\varphi$,
  $\varphi_1$, and $\varphi_2$ denote state formulas, and $\psi$,
  $\psi_1$, and $\psi_2$ denote path formulas, and $i$, $j$, and $k$
  are natural numbers:
\begin{align*}
  K,\pi \models \varphi  &\eqdef K,\pi(0) \models \varphi\\
  K,\pi \models \neg \psi &\eqdef K,\pi \not\models \psi\\                 
  K,\pi \models \psi_1 \land \psi_2
                 &\eqdef K,\pi\models \psi_1 \land K,\pi\models \psi_2\\
  K,\pi \models \psi_1 \lor \psi_2 
                 &\eqdef K,\pi \models \psi_1 \lor K,\pi\models \psi_2\\
  K,\pi \models X \psi   &\eqdef K,\pi_1 \models \psi\\
  K,\pi \models \psi_1 \U \psi_2 &\eqdef
     \E{j} K,\pi_j \models \psi_2 \land \A{0 \leq i < j} K,\pi_i \models \psi_1\\
  K,\pi \models \psi_1 \dU \psi_2 &\eqdef
      \A{j} K,\pi_j \not\models \psi_2 \limp \E{0 \leq i < j} K,\pi_i \models \psi_1\\
  K,\pi \models F \psi &\eqdef \E{j} K,\pi_j \models \psi\\
  K,\pi \models G \psi &\eqdef \A{j} K,\pi_j \models \psi\\[0.1in]
  K,s \models p &\eqdef p \in I(s)\\
  K,s \models \neg \varphi &\eqdef K,s \not\models \varphi\\                 
  K,s \models \varphi_1 \land \varphi_2
                 &\eqdef K,s\models \varphi_1 \land K,s\models \varphi_2\\
  K,s \models \varphi_1 \lor \varphi_2 
                 &\eqdef K,s \models \varphi_1 \lor K,s\models \varphi_2\\
  K,s \models A \varphi & \eqdef  \Land_{\pi \in \Pi^K_s} \pi \models \varphi \\
  K,s \models E \varphi & \eqdef  \Lor_{\pi \in \Pi^K_s} \pi \models \varphi
\end{align*}
\end{definition}
We say that $K$ satisfies $\varphi$ (or $\varphi$ holds in $K$),
denoted $K \models \varphi$, iff $\varphi$ holds in the designated
initial state: $K,s_0 \models \varphi$. %
For simplicity of presentation, we use sets of states as atomic propositions in temporal formulas, giving them the following interpretation: for a set of states $Y$,
\begin{equation*}
  K, s \models Y \eqdef s \in Y\,.
\end{equation*}

We write $\varphi[x]$ to indicate that the formula $\varphi$ may
contain an occurrence of $x$. An occurrence of $x$ in $\varphi$ is
\emph{positive} (or of \emph{positive polarity}) if $x$ occurs under
the scope of an even number of negations, and \emph{negative}
otherwise. For example, $p$ is positive in $\neg \EX \neg p$, and
negative in $\neg \EX p$. A subformula $x$ is \emph{pure} in $\varphi$
if all of its occurrences have the same polarity. For example, $p$ is
pure in $EF (p \land q \land EG p)$. We write $\varphi[x \subst y]$
for a formula obtained from $\varphi$ by replacing \emph{each
  occurrence} of $x$ by $y$.  This is equivalent to treating a formula
as a DAG with all common subformulas shared.

A formula $\varphi$ is \emph{universal} (i.e., in the language
\actlast{}) if all of its temporal path quantifiers are universal, and
is \emph{existential} (i.e., in the language \ectlast{}) if all of the
path quantifiers are existential.  In both cases, negation is only
allowed at the level of atomic propositions.  For example, $\AG (p
\limp AF q)$ is in \actlast{}, and $EF (p \land EG \neg q)$ is in
\ectlast{}. We extend this to subformulas as well and say that a
subformula is universal if it occurs only under the scope of universal
path quantifiers in negation normal form of the formula.

The fragment of \ctlast in which all formulas are of the form $A
\psi$, where $\psi$ is a path formula, is called \emph{Linear Temporal
  Logic} (\ltl{})~\cite{pnueli77}. The fragment in which every occurrence of a path
quantifier is immediately followed by a temporal operator is called
\emph{Computation Tree Logic} (\ctl{})~\cite{clarke81}. For example,
$\AG (p U q)$ is an \ltl formula, and $\AG A[p \U q]$ is a \ctl
formula. More details on temporal logic can be found
in~\cite{emerson90b,clarke99}.

\subsection{Simulation and Bisimulation}
\label{sec:simulation}
In this section, we review two property preserving relations between
\kripke structures: simulation and bisimulation.

\begin{definition}[Simulation]
  \label{def:simulation}\cite{milner71}
  Let $K = (AP, S, R, s_0,  I)$ and $K' = (AP', S', R', s'_0, I')$ be
  two \kripke structures and $X \subseteq (AP \cap AP')$ a set of
  common atomic proposition.  A relation $\rho \subseteq S \times S'$
  is a \emph{simulation} relation with respect to $X$ if and only if
  $\rho(s,s')$ implies that
  \begin{enumerate}
   \item $I'(s')\cap X = I(s) \cap X$, and
   \item $\A{t' \in S'} R'(s', t') \limp \E{t \in S} R(s,t) \land \rho(t,t')$.
  \end{enumerate}
\end{definition}

\noindent  A state $s$ simulates  a state $s'$ if
$(s, s') \in \rho$.  A \kripke structure $K$ \emph{simulates} $K'$ iff
the initial state of $K'$ is simulated by the initial state of $K$.
For example, $\mM$ in Figure~\ref{fig:xvariants} simulates $\mL$
in Figure~\ref{fig:k-one} via the relation
\begin{equation*}
  \rho_{\mM}^{\mL} = \{(b_0, a_0), (b_1, a_0)\}.
\end{equation*}

Intuitively, If $K$ simulates $K'$ then $K$ can match every behavior
of $K'$, i.e., the set of all behaviors of $K'$ is a subset of those
of $K$. Thus, if $K$ satisfies an \actlast formula, then so does $K'$.
\begin{theorem}\cite{browne88,grumberg94}
  \label{thm:simulation-actlast}
  Let $K$ and $K'$ be two \kripke structures such that $K$ simulates
  $K'$. Then, for any \actlast formula $\varphi$
  \begin{equation*}
    K \models \varphi \limp K' \models \varphi\,.
  \end{equation*}
\end{theorem}

A simulation relation whose inverse is also a simulation is called a
bisimulation:
\begin{definition}[Bisimulation]
  Let $K = (AP, S, R, s_0, I)$ and $K' = (AP', S', R', s'_0,  I')$ be
  two \kripke structures and $X \subseteq (AP \cap AP')$ a set of
  common atomic proposition. A relation $\rho \subseteq S \times S'$
  is a \emph{bisimulation} relation with respect to $X$ if and only if
  (a) $\rho$ is a simulation relation between $K$ and $K'$ with
  respect to $X$, and (b) $\rho^{-1} \subseteq S' \times S$ is a
  simulation relation between $K'$ and $K$ with respect to $X$.
\end{definition}

Two structures $K$ and $K'$ are \emph{bisimilar} iff there exists a
bisimulation relation $\rho$ that relates their initial states. We use
$\Bisim(K)$ to denote the set of all structures bisimilar to $K$ with
respect to all of the atomic propositions of $K$. For example, the
inverse of the relation $\rho_{\mM}^{\mL}$ above is a simulation as
well. Thus, $\mL$ and $\mM$ are bisimilar.

Intuitively, if $K$ and $K'$ are bisimilar, then they have equivalent
behaviors. The theorem below also indicates
that they satisfy the same temporal logic formulas.
\begin{theorem}\cite{browne88}
  \label{thm:bisimulation-ctlast}
  Let $K$ and $K'$ be two bisimilar \kripke structures. Then, for any
  \ctlast formula $\varphi$,
  \begin{equation*}
    K \models \varphi \liff K' \models \varphi\,.
  \end{equation*}
\end{theorem}

It is possible to extend the definition of bisimulation to
infinite-state models. Under such an interpretation, a computation
tree $T(K)$ of a \kripke structure $K$ is bisimilar to $K$.  This is
sufficient to show that a \ctlast formula cannot distinguish between a
\kripke structure and its tree unrolling, i.e., $K \models \varphi
\liff T(K) \models \varphi$. This fact is often used to give semantics
of \ctlast with respect to a computation tree of a \kripke structure
instead of with respect to the \kripke structure itself.  We say that
\ctlast is \emph{bisimulation closed}.  Note that not all temporal
logics share this property.  In particular, some quantified temporal
logics that are used in this article (see
Section~\ref{sec:quant-temp-logic}) are not bisimulation closed.

\subsection{Quantified Temporal Logic}
\label{sec:quant-temp-logic}

Quantified Temporal Logic (\qctlast) extends the syntax of \ctlast
with universal ($\forall$) and existential ($\exists$) quantifiers
over atomic propositions~\cite{kupferman97}.  For example, $\A{x} \EF
(x \limp EF (\neg x))$ is a \qctlast formula.  Here, we consider a
fragment in which only a single occurrence of a quantifier is allowed,
i.e.,
\begin{equation*}
\{\varphi, \A{x}\varphi, \E{x}\varphi \mid \varphi \in \ctlast\}.
\end{equation*}
For simplicity, we still call this fragment  \qctlast.

There are several different definitions of semantics of \qctlast with
respect to a \kripke structure; we consider three of these:
\emph{structure}~\cite{kupferman97}, \emph{tree}~\cite{kupferman97},
and \emph{bisimulation} which is introduced in~\cite{french01} under
the name \emph{amorphous}.

\heading{Structure Semantics.} 
Under structure semantics~\cite{kupferman97}, each bound variable $x$
is interpreted as a subset of the statespace. A universally quantified
formula $\A{x}\varphi$ is satisfied by $K$ under this semantics if
replacing $x$ by an arbitrary set always results in a formula that is
satisfied by $K$.

\begin{definition}[Structure Semantics]\cite{kupferman97}
  \label{def:qctl-structure}
  Let $K$ be a \kripke structure, and $\varphi$ a \ctlast formula.
  \emph{Structure semantics} of \qctlast, written $K \models_s
  \varphi$, is defined as follows:
  \begin{align*}
  K \models_s \varphi &\eqdef K \models \varphi\\
  K \models_s \A {x} \varphi  &\eqdef 
             \A{ Y \subseteq S}  K \models \varphi[x \subst Y]  \\
  K \models_s \E {x} \varphi  &\eqdef 
             \E{ Y \subseteq S}  K \models \varphi[x\subst Y]\;.
  \end{align*}

\end{definition}
That is, a formula $\A{x}\varphi[x]$ is satisfied by $K$ under
structure semantics if $\varphi[x]$ is true in $K$ under any
interpretation of the atomic proposition $x$.

An equivalent and more constructive definition can be given as well. Let
$K_{-x}$, pronounced ``$K$ minus $x$'', denote the result of removing
an atomic proposition $x$ from $K$. Formally,
\begin{equation*}
  K_{-x} \eqdef K \text{ with } AP_{-x} = AP \setminus \{x\}.
\end{equation*}
An \emph{$x$-variant} of a \kripke structure $K$ is a structure $K'$
such that $K'_{-x}$ is identical to $K$. For example, the set of all
$x$-variants of $\mL$ is shown in Figure~\ref{fig:k-one}.  A
formula $\A{x}\varphi[x]$ is satisfied by a \kripke structure $K$
under structure semantics if and only if $\varphi[x]$ is satisfied by
\emph{every} $x$-variant of $K$. This follows immediately from the
one-to-one correspondence between subsets of the statespace of $K$ and
labeling of $x$ in an $x$-variant.

We illustrate this semantics using the following formulas:
\begin{align*}
  P_1 &\eqdef \AG (x \limp \AX x);\\
  P_2 &\eqdef \AG ((\AX x) \lor (\AX \neg x));\\
  P_3 &\eqdef A ((X x) \lor (X \neg x)).
\end{align*}
$\mL \models_s \A{x}P_1$ since $P_1$ is satisfied by all $x$-variants
of $\mL$ (see Figure~\ref{fig:k-one}), but $\mM \not\models_s
\A{x}P_1$ since $P_2$ is not satisfied by the $x$-variant
$\variant{\mM}{4}$ of $\mM$ (see Figure~\ref{fig:xvariants}). The
results of evaluating the rest of the formulas on $\mL$ and $\mM$ are
summarized in the first three columns of Table~\ref{tbl:qctl-ex}.

\begin{figure}[t]
  \centering
  \renewcommand{\pspicgrid}{\pspicture}
\begin{pspicgrid}(0,-.5)(10,3.2)

\rput(.5,1){$\mM$}

\rput(4,1.7){$\variant{\mM}{1}$}
\rput(7,1.7){$\variant{\mM}{2}$}

\rput(4,-.3){$\variant{\mM}{3}$}
\rput(7,-.3){$\variant{\mM}{4}$}

\rput[bl](0,1.2){\input{fig-model}}

\rput[bl](4,0){\begin{pspicture}(0.5,0)(3,1.5)
\psset{framesep=0}
\cnodeput(1,0.5){a}{\begin{tabular}{@{}c@{}}$p$\\$ x$\end{tabular}}
\cnodeput(2.5,0.5){b}{\begin{tabular}{@{}c@{}}$p$\\$ x$\end{tabular}}

\psset{arrows=->}
\ncline{a}{b}
\nccircle[angleA=0]{->}{b}{.4}
\nccircle[angleA=0]{->}{a}{.4}
\end{pspicture}}

\rput[bl](4,2){\begin{pspicture}(0.5,0)(3,1.5)
\psset{framesep=0}
\cnodeput(1,0.5){a}{\begin{tabular}{@{}c@{}}$p$\\$\neg x$\end{tabular}}
\cnodeput(2.5,0.5){b}{\begin{tabular}{@{}c@{}}$p$\\$ \neg x$\end{tabular}}

\psset{arrows=->}
\ncline{a}{b}
\nccircle[angleA=0]{->}{b}{.4}
\nccircle[angleA=0]{->}{a}{.4}
\end{pspicture}}

\rput[bl](7,0){\begin{pspicture}(0.5,0)(3,1.5)
\psset{framesep=0}
\cnodeput(1,0.5){a}{\begin{tabular}{@{}c@{}}$p$\\$x$\end{tabular}}
\cnodeput(2.5,0.5){b}{\begin{tabular}{@{}c@{}}$p$\\$ \neg x$\end{tabular}}

\psset{arrows=->}
\ncline{a}{b}
\nccircle[angleA=0]{->}{b}{.4}
\nccircle[angleA=0]{->}{a}{.4}
\end{pspicture}}

\rput[bl](7,2){\begin{pspicture}(0.5,0)(3,1.5)
\psset{framesep=0}
\cnodeput(1,0.5){a}{\begin{tabular}{@{}c@{}}$p$\\$\neg x$\end{tabular}}
\cnodeput(2.5,0.5){b}{\begin{tabular}{@{}c@{}}$p$\\$x$\end{tabular}}

\psset{arrows=->}
\ncline{a}{b}
\nccircle[angleA=0]{->}{b}{.4}
\nccircle[angleA=0]{->}{a}{.4}
\end{pspicture}}

\end{pspicgrid}
  \caption{A \kripke structure $\mM$ and its $x$-variants: $\variant{\mM}{1}$, $\variant{\mM}{2}$, $\variant{\mM}{3}$, and $\variant{\mM}{4}$.}
  \label{fig:xvariants}
\end{figure}

\begin{table}[t]
  \centering
  \begin{tabular}{|c|c||c|c|c|}
    \hline
             & & 
                    \multicolumn{3}{c|}{\textbf{Quantification Semantics}}\\
    \textbf{Model} & \textbf{Property} & Structure & Tree   & Bisimulation\\
    \hline\hline
    $\mL$ & $\A{x}P_1$      & \true     & \false & \false \\
    $\mM$ & $\A{x}P_1$      & \false    & \false & \false \\[0.05in]
    $\mL$ & $\A{x}P_2$      & \true     & \true  & \false \\ 
    $\mM$ & $\A{x}P_2$      & \false    & \false & \false \\[0.05in]  
    $\mL$ & $\A{x}P_3$      & \true     & \true  & \true  \\
    $\mM$ & $\A{x}P_3$      & \true     & \true  & \true  \\\hline
  \end{tabular}
\vspace{0.1in}
  \caption{Satisfaction of QTL formulas  $\A{x}P_1$, $\A{x}P_2$, and $\A{x}P_3$ on
models $\mL$ and $\mM$ under different semantics of QTL.}
  \label{tbl:qctl-ex}

\vspace{-0.2in}
\end{table}

\heading{Tree Semantics.}
Under the tree semantics~\cite{kupferman97}, \qctlast formulas are
interpreted with respect to variants of a computation tree $T(K)$ of a
\kripke structure $K$.
\begin{definition}[Tree Semantics]~\cite{kupferman97}
  \label{def:qctl-tree}
  Let $K$ be a \kripke structure, and $\varphi$ a \ctlast formula.
  \emph{Tree semantics} of \qctlast, written $K \models_T \varphi$, is
  defined as follows:
  \begin{align*}
    K \models_T \varphi & \eqdef T(K) \models \varphi\\
    K \models_T \A{x}\varphi  &\eqdef  T(K) \models_s \A{ x }\varphi\\
    K \models_T \E{x}\varphi  &\eqdef  T(K) \models_s \E{ x }\varphi\;.
  \end{align*}
\end{definition}
That is, a formula $\A{x}\varphi[x]$ is satisfied by $K$ under tree
semantics if and only if it is satisfied by every $x$-variant of the
\emph{computation tree} of $K$. For example, $\mL \not\models_T
\A{x}P_1$ since $P_1$ is not satisfied by an $x$-variant
$\variant{T(\mL)}{1}$ of $T(\mL)$ shown in
Figure~\ref{fig:tree-unrolling}, and \mbox{$\mL \models_T \A{x}P_2$}
since every state in the tree unrolling $T(\mL)$ of $\mL$ has exactly
one successor. A few additional examples are given in the middle
column of Table~\ref{tbl:qctl-ex}.  We note that \qctlast under
structure and tree semantics is not bisimulation
closed~\cite{kupferman97}.

\heading{Bisimulation Semantics.}  
Prior to presenting bisimulation semantics, we need to introduce a notion
of $x$-bisimulation.  Let $K$ and $K'$ be two \kripke structures. The
structure $K'$ is \emph{$x$-bisimilar} to $K$ if and only if (a) the
atomic propositions $AP'$ of $K'$ extend atomic propositions $AP$ of
$K$ with a single atomic proposition $x$, i.e., $AP' = AP \cup \{x\}$,
and (b) $K'$ and $K$ are bisimilar with respect to $AP$. That is, $K'$
has exactly the same behaviors as $K$, except for the interpretation
of an additional atomic proposition $x$. For a \kripke structure $K$,
we use $\Bisim_x(K)$ to denote the set of all structures $x$-bisimilar
to $K$. For example, the $x$-variant $\variant{\mM}{4}$ of $\mM$ is
$\{x\}$-bisimilar to $\mM$. $\variant{\mM}{4}$ is also
$\{x\}$-bisimilar to $\mL$.  It is easy to observe that in general, the set $\Bisim_x(K)$ includes
all $x$-variants of the structure $K$, every structure bisimilar to
$K$, and every $x$-variant of a structure bisimilar to $K$.  The above
statement is included here just for clarity.

We are now ready to define bisimulation semantics. Under bisimulation
semantics, \qctlast formulas are interpreted with respect to
bisimulation variants of a \kripke structure.
\begin{definition}[Bisimulation (Amorphous) Semantics]\cite{french01}
  \label{def:qctl-bisimulation}
  Let $K$ be a \kripke structure, and $\varphi$ a \ctlast formula.
  \emph{Bisimulation semantics} of \qctlast, written $K \models_b
  \varphi$, is defined as follows:
  \begin{align*}
    K \models_b \varphi &\eqdef K \models \varphi\\
    K \models_b \A{ x} \varphi  &\eqdef  
         \A{K' \in \Bisim_x(K)} K' \models \varphi \\
    K \models_b \E{ x }\varphi  &\eqdef  
         \E{K' \in \Bisim_x(K) } K' \models \varphi\;.
   \end{align*}
\end{definition}
That is, a formula $\A{x}\varphi$ is satisfied by $K$ under bisimulation
semantics if and only if $\varphi$ is satisfied by every
$x$-bisimulation of $K$. For example, \mbox{$\mL \not\models_b
  \A{x}P_2$} since (a) $\mM$ is bisimilar to $\mL$, (b) any
$x$-variant of $\mM$ is $x$-bisimilar to $\mL$, and (c) $P_2$ is not
satisfied by the $x$-variant $\variant{\mM}{4}$ of $\mM$ (see
Figure~\ref{fig:xvariants}). On the other hand, $\mL \models_b
\A{x}P_3$ since $P_3$ is a temporal logic tautology, i.e., it is true
in any model. A few additional examples are given in the last column
of Table~\ref{tbl:qctl-ex}.

Note that each semantics extends the range of the interpretation of
the quantifiers. Thus, it is harder to satisfy a universal formula
under bisimulation semantics than under tree or structure semantics. The
following theorem formalizes the relationship between all three
\qctlast semantics, and is a corollary of a similar theorem proved by
French~\cite{french01}.

\begin{theorem}
  \label{thm:qctl-sem-reln}
  Let $\A{x} \varphi$ be a \qctlast formula, and $K$ a \kripke
  structure. Then, the following is true
  \begin{equation*}
    (K \models_b \A{x}\varphi) \limp (K \models_T \A{x}\varphi) \limp (K \models_s \A{x}\varphi)\;.
  \end{equation*}
  Furthermore, the implications are strict.
\end{theorem}
\begin{proof}
  The theorem follows from the fact that every tree unrolling of an
  $x$-variant of $K$ is an $x$-variant of $T(K)$ and that a tree
  unrolling $T(K)$ is bisimilar to $K$. Strictness of the first and
  the second implication is established by the examples in row~3 and
  row~1 of Table~\ref{tbl:qctl-ex}, respectively.
\end{proof}

\section{Towards Defining Vacuity}
\label{sec:vacuity-definition}
The first formal definition of vacuity is called \emph{propositional
antecedent failure} and was described by Beatty and
Bryant~\cite{beatty94}.  A formula of the form $\AG (p \limp q)$
suffers from antecedent failure on a model $K$ if its antecedent $p$
is not satisfiable in $K$. In particular, this means that the
consequent (or the right-hand side) of the implication does not effect
the validity of the formula.

\begin{figure}[t]
  \centering
  \let\pspicgrid=\pspicture
\begin{pspicgrid}(0,0)(11,2)

\rput[bl](0,0){%
  \begin{pspicture}(0,0)(2.5,2)%
      \cnodeput[framesep=3pt](1,1){a}{\begin{tabular}{@{}c@{}}$q$\end{tabular}}
      \cnodeput[framesep=1pt](2,1){b}{\begin{tabular}{@{}c@{}}$\neg q$\end{tabular}}
      \psset{arrows=->}
      \nccircle[angleA=0]{->}{a}{.3}
      \nccircle[angleA=0]{->}{b}{.3}
      \ncline{a}{b}
      \rput(.5,0.3){$\mN$}
    \end{pspicture}}

\rput[bl](3.5,0){%
  \begin{pspicture}(0,0)(2.5,2)%
      \cnodeput[framesep=0pt](1,1){a}{\begin{tabular}{@{}c@{}}$p$\\$q$\end{tabular}}
      \cnodeput[framesep=0pt](2.3,1){b}{\begin{tabular}{@{}c@{}}$p$\\$\neg q$\end{tabular}}
      \psset{arrows=->}
      \nccircle[angleA=0]{->}{a}{.3}
      \nccircle[angleA=0]{->}{b}{.3}
      \ncline{a}{b}
      \rput(.5,0.3){$\mO$}
    \end{pspicture}}

\rput[bl](7.5,0){%
  \begin{pspicture}(0,0)(2.5,2)%
      \cnodeput[framesep=0pt](1,1){a}{\begin{tabular}{@{}c@{}}$p$\\$\neg q$\end{tabular}}
      \cnodeput[framesep=0pt](2.3,1){b}{\begin{tabular}{@{}c@{}}$\neg p$\\$ q$\end{tabular}}
      \psset{arrows=->}
      \nccircle[angleA=0]{->}{b}{.3}
      \ncline{a}{b}

      \nput*{-45}{a}{$c_0$}
      \nput*{-45}{b}{$c_1$}
      \rput(.5,0.3){$\mP$}
    \end{pspicture}}

\end{pspicgrid}
  \caption{Sample models $\mN$, $\mO$, and  $\mP$. }
  \label{fig:samples}
\end{figure}

Beer et al.~\cite{beer01} have generalized antecedent failure to
arbitrary temporal formulas, calling the result \emph{temporal vacuity}.
Informally, if a formula $\varphi$ contains a subformula $\psi$ such that
replacing $\psi$ by any other formula does not affect the value of
$\varphi$, then $\varphi$ is vacuous in $\psi$.  Furthermore, \cite{beer01}
restricted vacuity to properties with  \emph{a single occurrence} of $\psi$.
We call this definition \emph{structural vacuity} and provide a formal
definition below:
\begin{definition}[Syntactic Vacuity]
  \label{def:formula-vacuity}
  \cite{beer97} A formula $\varphi$ in a temporal logic $L$ is
  \emph{syntactically vacuous} in a subformula $\psi$ (assuming
a single occurrence of $\psi$ in $\varphi$) in a model $K$ iff
  \begin{equation*}
    \A{\psi' \in L} K\models\varphi \liff K\models \varphi[\psi \subst \psi'].
  \end{equation*}
\end{definition}
When $\varphi$ is vacuous in $\psi$, we say $\varphi$ is
$\psi$-vacuous. A formula is vacuous if it is vacuous in any of its
subformulas. According to Definition~\ref{def:formula-vacuity},
non-vacuity of $\varphi$ with respect to a subformula $\psi$ is
witnessed by a formula $\varphi'$ of the form $\varphi' = \varphi[\psi
\subst \psi']$ for some $\psi' \in L$ such that $K \models \varphi$
and $K \not\models \varphi'$. For example, a non-vacuous satisfaction
of $\AG(r \limp \AF a)$ with respect to $\AF a$ can be witnessed by
falsification of $\AG(r \limp \false)$.

Definition~\ref{def:formula-vacuity} provides a useful
generalization of antecedent failure.  However, when Armoni et
al.~\cite{armoni03} attempted to generalize syntactic vacuity further (they called it
\emph{formula vacuity}),
to deal with multiple occurrences of subformulas, they found that
it has three major
weaknesses: (1) it makes vacuity of too many formulas debatable, (2)
it makes vacuity sensitive to changes in the model that do not (or
should not) affect the formula, and (3) it makes vacuity sensitive to
the syntax of the temporal logic.  We illustrate these weaknesses using
several examples inspired by (or sometimes taken directly from) Armoni et al.~\cite{armoni03}.

\heading{Weakness~1.}  
Consider the property
\begin{equation*}
  P_4 \eqdef \AG \left((\AX p) \lor (\AX \neg p)\right).
\end{equation*}
which means ``in every state, the next valuation of $p$ is computed
deterministically'', i.e., it is either true in all successors or
false in all successors.  This property can be vacuous in $AX p$ or
$AX \neg p$, since satisfaction of either disjunct is sufficient to
satisfy the entire property.  However, as we argue below, it should
never be vacuous in $p$ under any reasonable definition of vacuity.
Our reasoning is as follows. Take any \kripke structure $K$. Every
state of $K$ has at least one successor, and the proposition $p$ has
some value in each successor of every state. Thus, the value of $p$
directly influences the overall value of $P_4$. Hence, $P_4$ should
not be vacuous in $p$, in any \kripke structure.  However, according
to syntactic vacuity from Definition~\ref{def:formula-vacuity}, $P_4$
is $p$-vacuous in model $\mL$ in Figure~\ref{fig:k-one}, since $\mL$
satisfies $P_4$, $P_4[p \subst \true]$, and $P_4[p \subst \false]$.
This example shows that vacuity of some syntactically vacuous formulas
is debatable, and thus syntactic vacuity is not sufficiently strong.

\heading{Weakness~2.} 
Consider again the property $P_4$ defined above. We have already shown
that it is syntactically $p$-vacuous in $\mL$. Next, consider models
$\mN$, and parallel synchronous composition $\mO = \mL || \mN$ of
$\mL$ and $\mN$, both shown in Figure~\ref{fig:samples}. 
  The composition does not affect any of the original properties that
  were satisfied by $\mL$. However, it does affect the syntactic
  vacuity of $P_4$: $P_4$ is no longer syntactically $p$-vacuous in $\mO$.
  In particular, $\mO$ satisfies $P_4$ (just like $\mL$), but refutes
  \begin{equation*}
    P_4[p \subst q] = \AG \left((\AX q)  \lor (\AX \neg q)\right)\,.
  \end{equation*}
  Thus, composing $\mL$ with $\mN$ ``fixes'' syntactic vacuity of
  $P_4$, even though $\mN$ has no influence on satisfaction of $P_4$.
  This illustrates that syntactic vacuity is sensitive to
  ``irrelevant'' changes to the model.

  \heading{Weakness~3.}  Consider the property $P_5 \eqdef A (X q
  \limp XX q)$ and the model $\mP$ in Figure~\ref{fig:samples}. Assume
  that $P_5$ is interpreted in LTL. Since $\mP \models P_5[q \subst
  \psi]$ for any LTL formula $\psi$, $P_5$ is $q$-vacuous in $\mP$
  according to syntactic vacuity (see Definition~\ref{def:formula-vacuity}). 

  Let $X^{-1}$ denote the past operator meaning ``in the previous
  state''. Formally, $X^{-1} p$ is satisfied by a suffix $\pi_j$ of a
  path $\pi$ iff $j > 0$, and $p$ is satisfied by the suffix
  $\pi_{j-1}$. 

Let  LTL+P denote LTL  extended
  with the past operator.  Interpreted in LTL+P, $P_5$ is no longer
syntactically $q$-vacuous! The
  witness to non-vacuity is
\begin{align*}
  P_5[q \subst X^{-1} p] ={}& A ((X X^{-1} p) \limp (X X X^{-1} p)) \\
                        ={}& A (p \limp X p),
\end{align*}
which is falsified by $\mP$.  That is, syntactic vacuity of a formula
can change by re-interpreting the formula in a temporal logic with
more operators (without changing the formula itself),
allowing us to conclude that syntactic vacuity is
sensitive to the syntax of the logic with respect
to which the formula is defined.
 
In the rest of this section, we systematically develop a robust
definition of vacuity of temporal logic. We explore several semantic
definitions of vacuity starting with vacuity for propositional logic
and ending with a new definition of vacuity for temporal logic. We
argue that our definition is robust by showing that it is not affected
by non-essential changes to the model, nor by the number of available
logical operators. Note that unlike prior work~\cite{beer01,armoni03},
we do not distinguish between vacuity with respect to a particular
occurrence or several occurrences of a subformula. Instead, we present a
uniform treatment of the definition of vacuity that would allow the
user to make the distinction during use.
While we base the treatment below on subformula vacuity, all of our
results easily extend to vacuity with respect to arbitrary
subsets of occurrences.  Of course, when restricted to subformulas
with a single occurrence, all of the definitions of vacuity used in
this paper reduce to the original definition of Beer et al.~\cite{beer01}.

\subsection{Propositional Vacuity}
\label{sec:prop-vacu}
We start our exploration of vacuity with propositional logic.  A model
of a propositional formula $\varphi$ is just a boolean valuation of
all atomic propositions of $\varphi$. The value of $\varphi$ in a
model is a boolean value, either \true or \false.  Thus, we can check
the dependence of $\varphi$ on a subformula $\psi$ by checking whether
replacing $\psi$ by constants \true and \false affects the value of
$\varphi$. This leads to the following formal definition of
propositional vacuity.

\begin{definition}[Propositional Vacuity]
  \label{def:propositional-vacuity}
  A propositional formula $\varphi$ is \emph{vacuous} in a subformula
  $\psi$, or simply $\psi$-vacuous, in a model $K$ if and only if
  replacing $\psi$ by \true and \false does not affect the value of
  $\varphi$:
  \begin{equation*}
   (K \models \varphi[\psi \subst \true]) \liff (K \models \varphi[\psi \subst  \false])\;.
  \end{equation*}
\end{definition}
\noindent
A propositional formula is vacuous if it is vacuous in some subformula
$\psi$. Alternatively, vacuity of a propositional formula in a
model $K$ can be also expressed as validity of a quantified boolean formula
in $K$; that is, $\varphi$ is satisfied $\psi$-vacuously if and only if 
\begin{equation*}
  K \models \A{x}\varphi[\psi \subst x],
\end{equation*}
and $\varphi$ is falsified $\psi$-vacuously if and only if  
\begin{equation*}
  K \models \A{x}\neg \varphi[\psi \subst x]\;.
\end{equation*}
Propositional vacuity is robust for propositional formulas: vacuity of
a formula $\varphi$ is not affected by trivial changes to the model
(such as extending the model with new atomic propositions), nor by the
fragment of the propositional logic used to express $\varphi$.

One may conjecture that Definition~\ref{def:propositional-vacuity}
describes robust vacuity for temporal logic as well. However, this is
not the case.  For example, consider again the formula \[P_4 = \AG
\left((\AX p) \lor (\AX \neg p)\right)\]
According to our intuition discussed as part of Weakness 1 earlier
in this section, $P_4$ should not be satisfied $p$-vacuously.
Yet, in any model,
\[\begin{array}{lclcl}
P_4[p \subst \true] & = & \AG \left((\AX  \true) \lor
  (\AX \neg \true)\right) & = & \true, \mbox{ and}\\
P_4[p \subst \false] & = & \AG \left((\AX \false) \lor (\AX \neg
  \false)\right) & = & \true.
\end{array}\]
Thus, by Definition~\ref{def:propositional-vacuity},
$\varphi$ is $p$-vacuous.

\subsection{Structure Vacuity}
\label{sec:structure-vacuity}

Proposition vacuity interprets a model as a mapping from 
\emph{every} state of the model to boolean values \true and \false.
This is a limitation when trying to extend this definition to temporal formulas:
replacing a subformula only by
the constants \true and \false is not sufficient for identifying
whether the subformula is important.  Following this observation, we
extend the definition of vacuity to account for all subsets of the
statespace $S$. The resulting definition, originally introduced
in~\cite{armoni03} under the name \emph{structure} vacuity, is given
below.

\begin{definition}[Structure Vacuity]\cite{armoni03}
  \label{def:structure-vacuity}
  A temporal logic formula $\varphi$ is \emph{structure}
  $\psi$-vacuous in a model $K$ if and only if either
  \begin{equation*}
    \A{Y \subseteq S} K \models \varphi [\psi \subst Y]\,,\text{ or}
   \end{equation*}
   \begin{equation*}
     \A{Y \subseteq S} K \models \neg \varphi[\psi \subst Y]\,,
   \end{equation*}
   where $S$ is the statespace of $K$.
\end{definition}
Alternatively, structure vacuity can be expressed as validity of a
quantified temporal logic formula under structure semantics; that is,
$\varphi$ is satisfied structure $\psi$-vacuously if and only if
\begin{equation*}
  K \models_s \A{x}\varphi[\psi \subst x]\,,
\end{equation*}
and $\varphi$ is falsified structure $\psi$-vacuously if and only if
\begin{equation*}
  K \models_s \A{x}\neg \varphi[\psi \subst x]\,.
\end{equation*}

Definition~\ref{def:structure-vacuity} makes vacuity too dependent on
a particular model of the system. This leads to undesired
side-effects. For example, consider again the property $P_4 = \AG
\left( (\AX p) \lor (\AX \neg p) \right)$ and models $\mL$ and $\mM$
from Figure~\ref{fig:k-one} and Figure~\ref{fig:xvariants},
respectively.  The two models are bisimilar and cannot be
distinguished by any temporal logic formula. However, recall that
according to Definition~\ref{def:structure-vacuity}, $P_4$ is
$p$-vacuous in $\mL$, and yet it is not $p$-vacuous in $\mM$. Thus,
structure vacuity is not robust for temporal logic.

\subsection{Bisimulation Vacuity}
\label{sec:bisimulation-vacuity}
The example in Section~\ref{sec:structure-vacuity} illustrates that it is not
sufficient to define vacuity with respect to a single
\emph{particular} model $K$. Instead, a robust definition of vacuity
must also take into account any model that is behaviorally
equivalent to $K$.  For temporal logic, two models are considered to
be behaviorally equivalent if and only if they are bisimilar. This
leads to the following, robust, definition of vacuity.

\begin{definition}[Bisimulation Vacuity]
  \label{def:vacuity}
  A temporal logic formula $\varphi$ is \emph{bisimulation}
  $\psi$-vacuous in a \kripke structure $K$ if and only if it is
  structure $\psi$-vacuous both in $K$ and in every structure
  bisimilar to $K$.  That is, either
  \begin{equation*}
    \A{K' \in
      \Bisim(K)}\A{Y \subseteq S'} K' \models \varphi [\psi \subst
    Y]\,, \text{ or}
  \end{equation*}
  \begin{equation*}
    \A{K' \in \Bisim(K)}\A{Y \subseteq S'} K'\models \neg \varphi [\psi \subst Y]\,,
  \end{equation*}
  where $S'$ denotes the statespace of $K'$.
\end{definition}
Alternatively, structure vacuity can be expressed as validity of a
quantified temporal logic formula under bisimulation semantics; that is,
$\varphi$ is satisfied bisimulation $\psi$-vacuously if and only if
\begin{equation*}
  K \models_b \A{x}\varphi[\psi \subst x]\,,
\end{equation*}
and $\varphi$ is falsified bisimulation $\psi$-vacuously if and only if 
\begin{equation*}
  K \models_b \A{x}\neg \varphi[\psi \subst x]\,.
\end{equation*}
That is, $\varphi[\psi \subst x]$ is either satisfied or violated in
every model that is $x$-bisimilar to $K$. For example, the property
$P_4$ is not bisimulation vacuous in either $\mL$ or $\mM$.

In the next section, we describe some of the key properties of
bisimulation vacuity and argue that it provides a uniform definition of
robust vacuity for both linear and branching time logics. 

\subsection{Properties of Bisimulation Vacuity}
\label{sec:prop-amorph-vacu}

  For \ctlast, bisimulation vacuity is more strict than either
  structure or syntactic vacuity, i.e., if a formula is vacuous w.r.t.
  bisimulation vacuity, then it is vacuous w.r.t. to structure and
  syntactic definitions of vacuity as well, but the converse is not
  true in general.

\begin{theorem}
  \label{thm:bism-vacuity-strong}
  Let $K$ be a \kripke structure, $\varphi$ be an \actlast formula, and
  $\psi$ be a subformula of $\varphi$. Then, if $\varphi$ is bisimulation
  vacuous in $\psi$ (in $K$) then (a) $\varphi$ is structure vacuous in
  $\psi$,  and (b) $\varphi$ is syntactically vacuous in $\psi$
  w.r.t. \ctlast.
\end{theorem}
\begin{proof}
  Part (a) is a direct consequence of Theorem~\ref{thm:qctl-sem-reln}.
  
  To prove part (b), we show that for \ctlast, structure vacuity
  implies syntactic vacuity. Let $K = (AP, S, R, s_0, I)$ be a \kripke
  structure. By Definition~\ref{def:formula-vacuity}, $\varphi$ is
  syntactically vacuous in $\psi$ iff for any \ctlast formula $\psi'$,
  $K \models \varphi$ iff $K \models \varphi[\psi \subst \psi']$. Note
  that $\psi'$ is a state formula. Let $Y$ be the set of all states
  that satisfy $\psi'$. Formally, $Y = \{s \in S | K,s \models
  \psi'\}$. Then $K \models \varphi[\psi \subst \psi']$ iff $K \models
  \varphi[\psi \subst Y]$. Thus, for \ctlast, structure vacuity is
  more strict than syntactic vacuity: if $\varphi$ is structure
  vacuous in $\psi$, then $\varphi$ is syntactically vacuous in
  $\psi$.
\end{proof}

  In the rest of this section, we show that while bisimulation vacuity
  is \emph{not} too strict, i.e., it does capture the ``obvious''
  cases of vacuity, it is strict enough to be robust, i.e., it does not
  suffer from the three weaknesses identified in the beginning of this
  section.

Temporal logic tautologies are the most obvious examples of vacuous
formulas. We show that they are vacuous under bisimulation vacuity.
\begin{proposition}
  \label{thm:tautology-vacuity}
  Let $\varphi$ be a temporal logic formula with at least one atomic
  proposition, say $p$. If $\varphi$ is either valid or unsatisfiable, then it
  is bisimulation $p$-vacuous in any model.
\end{proposition}
\begin{proof}
  The theorem follows from the fact that validity is invariant under
  substitution of atomic propositions with fresh variables. That is,
  if $\varphi[p]$ is a valid formula with a proper subformula $p$, and
  $x$ is an atomic proposition that does not occur in $\varphi$, then
  $\varphi[p \subst x]$ is valid as well.
\end{proof}

For example, consider the property
\begin{equation*}
P_6 = (\EX p) \lor (\AX \neg p).
\end{equation*}
Replacing $p$ by $x$ in $P_6$ yields
\begin{equation*}
  P_6[p \subst x] = (\EX x) \lor (\AX \neg x),
\end{equation*}
which is a tautology. Hence, $\A{x} P_6[p\subst x]$ is satisfied by
any model under any semantics of \qctlast from
Section~\ref{sec:quant-temp-logic}.  Thus, property $P_6$ is
bisimulation $p$-vacuous in any model.

Bisimulation vacuity is able to detect vacuity even if the formula itself
is not a tautology, but contains a non-trivial tautology as a proper
subformula.  This follows from the proof of
Proposition~\ref{thm:tautology-vacuity}.
\begin{corollary}
  \label{thm:with-tautology-vacuity}
  Let $\varphi$ be a temporal logic formula, and $\psi$ be a proper
  non-constant subformula of $\varphi$ with an atomic proposition
  $p$. If $\psi$ is either valid or unsatisfiable and $\varphi$ does
  not contain $p$ outside of $\psi$, then $\varphi$ is
  $p$-bisimulation vacuous in any model.
\end{corollary}

For example, consider the property 
\begin{equation*}
  P_7 = \AG (q \land ((\EX p) \lor (\AX \neg p))).
\end{equation*}
Since a tautology can always be replaced by a constant, $P_7$ is
equivalent to
\begin{align*}
  &   \AG (q \land ((\EX p) \lor (\AX \neg p))) \\
={}&   \AG (q \land \true)\\
={}&   \AG (q).
\end{align*}
Hence, $P_7$ does not depend on $p$ and is $p$-vacuous in any model.
Note that since bisimulation vacuity is stricter than either structure
or syntactic vacuity, both Proposition~\ref{thm:tautology-vacuity} and
Corollary~\ref{thm:with-tautology-vacuity} extend to structure and
syntactic vacuity as well.

Bisimulation vacuity is strict enough to exclude vacuity that can be
``fixed'' by non-essential changes to the model. In particular, it can
distinguish between two models only if temporal logic can distinguish
between them as well. Thus, two models that agree on all temporal
logic formulas, also agree on their
bisimulation vacuity. 
\begin{proposition}
  \label{thm:bisimilarity-and-vacuity}
  Let $\varphi$ be a temporal logic formula, $\psi$ be a subformula of
  $\varphi$, and $K$ and $K'$ be two bisimilar \kripke structures.
  Then, $\varphi$ is $\psi$-vacuous in $K$ iff it is $\psi$-vacuous in
  $K'$.
\end{proposition}
\begin{proof}
  The proof follows immediately from the definition of bisimulation
  vacuity.
\end{proof}

For example, the model $\mL$ in Figure~\ref{fig:k-one} and the model
$\mM$ in Figure~\ref{fig:xvariants} are bisimilar. Thus, they agree on
satisfaction and vacuity of all temporal logic formulas.  In
particular, property $P_4$ (see Weakness~1) is not bisimulation
$p$-vacuous in either model.

An important consequence of
Proposition~\ref{thm:bisimilarity-and-vacuity} is that bisimulation
vacuity is not affected by parallel synchronous composition. That is,
if a formula is vacuous with respect to a component, then it is
vacuous with respect to the whole system as well.
\begin{corollary}
  \label{thm:composition-and-vacuity}
  Let $\varphi$ be a temporal logic formula, $\psi$ be a subformula of
  $\varphi$, and $K$ and $K'$ be two \kripke structures. If $\varphi$
  is bisimulation $\psi$-vacuous in $K$, then it is bisimulation
  $\psi$-vacuous in the parallel synchronous composition $K || K'$.
\end{corollary}
\begin{proof}
  This follows from the fact that $K$ and $K || K'$ are bisimilar with
  respect to atomic propositions of $K$.  For $K = (AP, S, R, s_0,
  I)$ and $K' = (AP', S', R', s'_0, I')$, let $K || K' = (AP \cup AP', S
  \times S',  R_{||}, (s_0, s'_0), I_{||})$ be their parallel synchronous
  composition (see Definition~\ref{def:sync-composition}).
  Then, the relation
  \begin{equation*}
    \rho \eqdef \{ (s, (s, t)) \mid s \in S, t \in S' \}
  \end{equation*}
  is a bisimulation between $K$ and $K || K'$.
\end{proof}

For example, consider again the example given in Weakness~2. The
formula $P_4$ is not bisimulation vacuous in the model $\mL$
(Figure~\ref{fig:k-one}), and its vacuity status does not change when
$\mL$ is composed with $\mN$ (Figure~\ref{fig:samples}); nor does its
vacuity status change when $\mL$ is composed with any other model that
does not affect the satisfaction of $P_4$.

In summary, we argue that bisimulation vacuity is robust and does not
suffer from the three weaknesses described in the beginning of this
section. Bisimulation vacuity is stricter than syntactic vacuity -- it
considers less formulas to be vacuous (Weakness~1). It is invariant
under bisimulation and cannot be affected by changes of the model that
are ``irrelevant'' to a property being checked (Weakness~2). Finally,
it is defined on the semantics of the temporal logic and, hence, is
independent of the syntax (Weakness~3). At the same time, it agrees
with syntactic vacuity (and other similar definitions) in all of the
``obvious'' cases of vacuity.

\section{Complexity of Vacuity Detection}
\label{sec:compl-vacu-detect}

In this section, we present algorithms for bisimulation vacuity
detection and analyze their complexity. We show that in general, the
complexity of bisimulation vacuity detection of a branching-time logic
is the same as the complexity of the satisfiability problem for that
logic. We then explore several practically important fragments of
branching time logics. We show that the complexity of bisimulation
vacuity detection for those fragments is in the same complexity class
as model-checking. In the rest of the article, we use the terms
``vacuity'' or ``robust vacuity'' to mean ``bisimulation vacuity'',
unless stated otherwise.

\subsection{Complexity of Detecting Bisimulation Vacuity}
\label{sec:compl-detect-amorhp}
We begin our study of complexity of detecting vacuity for branching time
logics with an example.
Let $\varphi$ be a
temporal logic formula over a single atomic proposition $p$.   That is,  while
there might be several occurrences of $p$ in $\varphi$, no other
atomic proposition is allowed.
Now consider the problem of detecting vacuity of $\varphi$
  with respect to model $\mL$ from Figure~\ref{fig:k-one}.  Note that
  every \kripke structure with a single atomic proposition $x$ is
  $p$-bisimilar to some \kripke structure in $\Bisim_x(\mL)$. Thus,
  $\varphi$ is satisfied $p$-vacuously by $\mL$ iff $\varphi[p \subst
  x]$ is a tautology. Similarly, $\varphi$ is falsified $p$-vacuously
  by $\mL$ iff $\varphi[p \subst x]$ is unsatisfiable. Thus, the
  problems of validity and satisfiability of $\varphi$ are reduced to
  detecting vacuity of $\varphi$ with respect to $\mL$.
We use this example as an intuition for formulating and proving the general
complexity result:
\begin{theorem}
  \label{thm:vacuity-complexity}
  Deciding whether a formula $\varphi$ is bisimulation $\psi$-vacuous
  is EXPTIME-complete for \ctl, and 2EXPTIME-complete for \ctlast.
  \end{theorem}
\begin{proof}
  To proof completeness, we need to show (1) membership and (2)
  hardness To show membership, we reduce bisimulation vacuity to
  model-checking quantified temporal logic under tree semantics. To
  show hardness, we reduce temporal logic satisfiability to
  bisimulation vacuity.

  \emph{Membership.}  Recall that detecting bisimulation vacuity is
  reducible to model-checking a quantified temporal logic formula
  under bisimulation semantics (see
  Section~\ref{sec:bisimulation-vacuity}).  Here, we reduce
  model-checking under bisimulation semantics to model-checking under
  tree semantics, which was shown by Kupferman in \cite{kupferman97}
to be in EXPTIME for E\qctl and in 2EXPTIME for
E\qctlast.

Formally, let $K = (AP, S, R, s_0, I)$ be a \kripke structure, and $m$
be a natural number. We define a \kripke structure $K^m$ to be the
tuple
  \[(AP, S \times [0, (m-1)], R^m, \tuple{s_0, 0}, I^m)\;,\]
  where the transition relation and the labeling function are defined
  as follows:
  \begin{align*}
    (\tuple{s,i}, \tuple{t,j}) \in R^m &\liff (s, t) \in R\\
    I^m(\tuple{s,i}) &\eqdef I(s)\;.
  \end{align*}
  Intuitively, $K^m$ is the result of duplicating each successor of
  $K$ $m$ times.

  Let $\E{x}\varphi$ be an E\qctl formula. We show that $K$ satisfies
  $\E{x}\varphi$ under bisimulation semantics iff $K^{|\varphi|}$
  satisfies $\E{x}\varphi$ under tree semantics, i.e.,
  \[
  K \models_b \E{x}\varphi \liff K^{|\varphi|} \models_T \E{x}\varphi\,.
  \]

  The proof of the ``if'' direction is trivial since $K^{m}$ is
  bisimilar to $K$ for any $m$.

  The proof of the ``only if'' direction uses the proof of the small
  model theorem for \ctl~(Theorem~6.14 in~\cite{emerson90b}). Assume
  that $K \models_b \E{x}\varphi$. Then there exists a computation
  $T$ such that (a) $T$ is bisimilar to $K$, and (b) $T$ satisfies
  $\varphi$ with respect to structure semantics, i.e., $T \models_s
  \varphi$. By the proof of the small model theorem for \ctl~(see
  proof of Theorem~6.14 in~\cite{emerson90b}), there exists a subtree
  $T'$ of $T$ such that
  \begin{enumerate}
  \item $T' \models_s \varphi$;
  \item $T'$ is bisimilar to $K$;
  \item the branching degree of $T'$ is bounded by $d_K + |\varphi|$,
    where $d_K$ is the branching degree of $K$.
  \end{enumerate}

  Let $T^{|\varphi|}$ be the computation tree of
  $K^{|\varphi|}$. Since $K^{|\varphi|}$ is bisimilar to $K$, by
  transitivity, $T^{|\varphi|}$ is bisimilar to $T'$. Furthermore, the
  branching degree of $T^{|\varphi|}$ is greater or equal to the
  branching degree of $T'$.  Hence, $T'$ is a subtree of
  $T^{|\varphi|}$. Therefore, $T^{|\varphi|} \models_s \varphi$ and
  $K^{|\varphi|} \models_b \E{x}\varphi$.

  The proof for \qctlast is based on the equivalent Small Model
  Theorem for \ctlast (Theorem~3.2 in~\cite{emerson84b}) and is
  otherwise identical to the one above.

  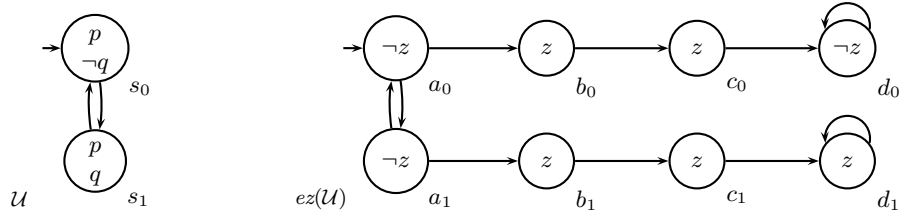
\begin{figure}[t]
    \centering
    \begin{pspicture}(0,0)(9,3)

  \rput[bl](-1,0.5){%
  \begin{pspicture}(0,0)(2,3)
    \small
    \rput(0.3,2){\pnode{init}}

    \cnodeput[framesep=0pt](1,2){a}%
    {\begin{tabular}{@{}c@{}}$p$\\$\neg q$\end{tabular}}
    \cnodeput[framesep=0pt](1,0.5){b}%
    {\begin{tabular}{@{}c@{}}$p$\\$q$\end{tabular}}
    
    \psset{arrows=->}
    \ncline{init}{a}

    \ncarc{a}{b}
    \ncarc{b}{a}
    
    \nput*{-45}{a}{$s_0$}
    \nput*{-45}{b}{$s_1$}

    {\normalsize \rput(0,0){$\mU$}}
  \end{pspicture}}

  \rput[bl](3,0.5){%
  \begin{pspicture}(0,0)(5,3)
    \small

    \rput(0.3,2){\pnode{init}}

    \cnodeput[framesep=4pt](1,2){a0}%
    {\begin{tabular}{@{}c@{}}$\neg z$\end{tabular}}
    \cnodeput[framesep=4pt](1,0.5){a1}%
    {\begin{tabular}{@{}c@{}}$\neg z$\end{tabular}}

    \cnodeput[framesep=4pt](3,2){b0}%
    {\begin{tabular}{@{}c@{}}$z$\end{tabular}}
    \cnodeput[framesep=4pt](5,2){c0}%
    {\begin{tabular}{@{}c@{}}$z$\end{tabular}}
    \cnodeput[framesep=2.5pt](7,2){d0}%
    {\begin{tabular}{@{}c@{}}$\neg z$\end{tabular}}

    \cnodeput[framesep=4pt](3,0.5){b1}%
    {\begin{tabular}{@{}c@{}}$z$\end{tabular}}
    \cnodeput[framesep=4pt](5,.5){c1}%
    {\begin{tabular}{@{}c@{}}$z$\end{tabular}}
    \cnodeput[framesep=4pt](7,.5){d1}%
    {\begin{tabular}{@{}c@{}}$z$\end{tabular}}

    \psset{arrows=->}
    \ncline{init}{a0}

    \ncarc{a0}{a1}
    \ncarc{a1}{a0}

    \ncline{a0}{b0}
    \ncline{b0}{c0}
    \ncline{c0}{d0}
    \nccircle[angleA=0]{->}{d0}{.3}    

    \ncline{a1}{b1}
    \ncline{b1}{c1}
    \ncline{c1}{d1}
    \nccircle[angleA=0]{->}{d1}{.3}

    \nput*{-45}{a0}{$a_0$}
    \nput*{-45}{a1}{$a_1$}
    \nput*{-45}{b0}{$b_0$}
    \nput*{-45}{b1}{$b_1$}
    \nput*{-45}{c0}{$c_0$}
    \nput*{-45}{c1}{$c_1$}
    \nput*{-45}{d0}{$d_0$}
    \nput*{-45}{d1}{$d_1$}

    {\normalsize \rput(0,0){$\ez(\mU)$}}

  \end{pspicture}} 

\end{pspicture}
    \caption{A \kripke structure $\mU$ with atomic propositions $p$ and $q$, and  its encoding $\ez(\mU)$ using a single atomic proposition $z$.}
    \label{fig:ez-example}
  \end{figure}
  
  \emph{Hardness for \ctl.} We have already shown that deciding
  satisfiability of a \ctl formula with a single atomic proposition is
  reducible to detecting bisimulation vacuity. Now, we reduce
  satisfiability of \ctl to satisfiability of a \ctl formula
  restricted to a single atomic proposition.  The idea is to encode
  the atomic propositions labeling each state by a structure attached
  to that state. For example, consider a model $\mU$ with atomic
  propositions $p$ and $q$ and its encoding, $\ez(\mU)$, that uses a
  single atomic proposition $z$ (see
  Figure~\ref{fig:ez-example}). States $a_0$ and $a_1$ of $\ez(\mU)$
  correspond to states $s_0$ and $s_1$ of $\mU$, respectively.  The
  structure rooted at $b_0$ encodes the labeling of the atomic
  propositions at $s_0$: state $b_0$ is labeled with $z$ to indicate
  that it is the root of the encoding structure; state $c_0$ is
  labeled with $z$ to indicate that $s_0$ is labeled with $p$, and
  $d_0$ is labeled with $\neg z$ to indicate that $s_0$ is labeled
  with $\neg q$. Similarly, the structure rooted at $b_1$ encodes the
  labeling of the atomic propositions at $s_1$.
  
  Formally, let $K = (AP, S, R, s_0, I)$ be a \kripke structure, $n =
  |AP|$ denote the number of atomic propositions, and $o:AP \to [0,n]$
  be some enumeration of atomic propositions. Then \kripke structure
  $\ez(K)$ is the tuple
  \[
  (\{z\}, S \times [0,(n+1)], R_{\ez}, \tuple{s_0, 0},  I_{\ez})\,,
  \]
  where $z$ is a new atomic proposition not in $AP$, and the transition
  relation and the labeling function are defined as follows:
  \begin{align*}
  (\tuple{s,i}, \tuple{t,j}) \in R_{\ez} &\liff \begin{cases}
    (s,t) \in R & \text{if } i = j = 0\\
    \true       & \text{if } s = t \land (j = i + 1 \lor i = j = n + 1)\\
    \false      & \text{otherwise}
    \end{cases}\\
    I_{\ez}(\tuple{s,i}) &\eqdef \begin{cases}
      \setof{\neg z} & \text{if } i = 0\\
      \setof{z}      & \text{if } i = 1\\
      \setof{z}      & \text{if } \E{p \in AP} 
                                     o(p) = (i - 1) \land p \in I(s)\\
     \setof{\neg z}  & \text{otherwise}                                     
      \end{cases}
  \end{align*}
  Given a \ctl formula $\psi$ with atomic propositions in $AP$, we
  replace each atomic proposition with a temporal logic formula over a
  new atomic proposition $z$.
  For example, the formula $E[\true \U
  \neg p \land A[\false \dU q]]$ is translated into 
  \begin{align*}
  E[\neg z &\U (\EX z) \land \AX (z \limp \AX \neg z) \land{} \\
           &\phantom{{}\U{}}  (\EG \neg z) \land 
              A[z \dU \neg z \limp ((\EX z) \land \AX (z \limp \AX\AX z))]]\,.
  \end{align*}              
  The translation increases the size of $\psi$ by a factor of
  $|\psi|$ due to the extra $\AX$ operators. 

Formally, we define a
  function $f$ such for any \ctl formula $\psi$, the following
conditions hold: (a) $f(\psi)$ only
  contains one atomic proposition, $z$, and (b) $f(\psi)$ and $\psi$ are
  equisatisfiable. We define $f$ by induction on the structure of
  $\psi$, showing just the ``interesting'' cases ($f$ distributes over
the operators in other cases).
  \begin{align*}
    f(p) &\eqdef (\EX z) \land AX (z \limp \AX^{o(p)+1} z)\\
    f(\neg p) &\eqdef (\EX z) \land AX (z \limp \AX^{o(p)+1} \neg z)\\
    f(\EX \psi_1) &\eqdef \EX (\neg z \land f(\psi_1))\\
    f(\AX \psi_1) &\eqdef \EX (\neg z) \land \AX (\neg z \limp f(\psi_1))\\
    f(E[\psi_1 \U \psi_2]) &\eqdef 
                  E[\neg z \land f(\psi_1) \U \neg z \land f(\psi_2)]\\
    f(A[\psi_1 \U \psi_2]) &\eqdef EG (\neg z) \land 
                  A[\neg z \limp f(\psi_1) \U \neg z \limp f(\psi_2)]\\
    f(E[\psi_1 \dU \psi_2]) &\eqdef 
                  E[\neg z \land f(\psi_1) \dU \neg z \land f(\psi_2)]\\
    f(A[\psi_1 \dU \psi_2]) &\eqdef EG (\neg z) \land 
                  A[\neg z \limp f(\psi_1) \dU \neg z \limp f(\psi_2)]
   \end{align*}
   
Since model $K$ satisfies a property $\psi$,  $\ez(K)$ satisfies
   $f(\psi)$. 

\vskip 0.1in

For the other direction, let $M = (\{z\}, S_M, R_M, s_0^M, I_M)$ be a
model for $f(\psi)$. Let $S_K$ be the smallest subset of $S_M$ that
satisfies the following two conditions:
\begin{gather*}
  \{s_0^M\} \in S_K,\text{ and} \\
  \A{s \in S_K} \{t \in S_M \mid I_M(t) = \{\neg z\} \land (s,t) \in R_M\} \subseteq S_K.
\end{gather*}
That is, $S_K$ includes the initial states and all states labeled with
$\neg z$ that are reachable from the initial state by other states
labeled with $\neg z$.

Let $K = (AP, S_K, R_K, s_0^M, I_K)$, where
\begin{align*}
  (s,t) \in R_K &\liff (s,t) \in R_M\\
  \{p\} \in I_K(s) &\liff M,s \models AX (z \limp \AX^{o(p)+1} z)\\
  \{\neg p\} \in I_K(s) &\liff M,s \models AX (z \limp \AX^{o(p)+1} \neg z).
\end{align*}
Then $K$ is a model for $\psi$: $K \models \psi$.  Note that
universal path quantifiers in the encoding of propositions (i.e.,
$f(p)$ and $f(\neg p)$ given above) ensure that the labeling $I_K$ is
consistent (i.e., no state is labeled with both $p$ and $\neg p$). The
existential path quantifiers in this encoding ensure that the
transition relation of $K$ is total.

   Since \ctl satisfiability has been shown to be
   EXPTIME-hard~\cite{fischer79}, this gives us the desired result.

   \emph{Hardness for \ctlast.} As in the proof of hardness for \ctl,
   we reduce satisfiability of \ctlast to satisfiability of a \ctlast
   formula restricted to a single atomic proposition.  For the models,
   we use the same encoding as for \ctl.
   
   To translate formulas, we define a function $g$ such for any
   \ctlast formula $\psi$, the following conditions hold: (a) $g(\psi)$ 
contains only one atomic
   proposition, $z$, and (b) $g(\psi)$ and $\psi$ are equisatisfiable.
   We define $g$ by induction on the structure of $\psi$, again showing
just the ``interesting'' cases.
  \begin{align*}
    g(p) &\eqdef (\EX z) \land AX (z \limp X^{o(p)+1} z)\\
    g(\neg p) &\eqdef (\EX z) \land AX (z \limp X^{o(p)+1} \neg z)\\
    g (E \psi_1) &\eqdef E ((G \neg z) \land g(\psi_1))\\
    g (A \psi_1) &\eqdef (EG \neg z) \land A ((G\neg z) \limp g(\psi_1))
   \end{align*}
   The rest of the proof proceeds the same way as for \ctl.  This
   establishes hardness in 2EXPTIME since \ctlast satisfiability has
   been shown to be 2EXPTIME-hard~\cite{vardi85}.
\end{proof}

Theorem~\ref{thm:vacuity-complexity}
suggests that bisimulation vacuity detection for
\ctlast and even for  \ctl is not computationally tractable.
However, we show that there are several important fragments of \ctlast
for which vacuity detection is in the same complexity class as
model-checking, and thus is tractable.  We study these in
the rest of this section, starting with monotone
formulas and continuing with \actlast and \ectlast.

\subsection{Vacuity and Monotone Formulas}
\label{sec:vacu-monot-form}
In this section, we study the problem of vacuity detection for
monotone formulas. We make two contributions. First, we show that 
vacuity detection for monotone formulas 
is reducible to model-checking. Our
algorithm is a natural extension of the vacuity detection algorithms
of Beer et al.~\cite{beer01} and Kupferman and
Vardi~\cite{kupferman99}. Second, we show that detecting whether a
formula expressed in a given temporal logic
is monotone is as hard as deciding the
satisfiability problem for this logic. This means that simple
monotonicity checks, such as restricting vacuity to a single
occurrence as in \cite{beer01}, or relying on polarity of occurrences,
as in \cite{armoni03}, can not be cheaply extended to the full temporal
logic.

\begin{definition}[Monotone Formula]
  A formula $\varphi$ is \emph{monotonically increasing} in a
  subformula $\psi$ if whenever $(x \limp y)$ is valid, so is
  $(\varphi[\psi \subst x] \limp \varphi[\psi \subst y])$.  It is
  \emph{monotonically decreasing} in $\psi$ if whenever $(x \limp y)$
  is valid, so is $(\varphi[\psi \subst x] \lpmi \varphi[\psi \subst
  y])$.  We say that $\varphi$ is \emph{monotone} in $\psi$ if it is
  either monotonically increasing or monotonically decreasing in
  $\psi$.
\end{definition}
For example, the formula $\AG (p \limp \AF q)$ is monotonically
decreasing in $p$ and is monotonically increasing in $q$; the formula
$\AG (p \land \neg p)$ is monotone in $p$, and the formula $\AG (p \limp
\AF (p \land q))$ is not monotone in $p$.

\begin{figure}[t]

  \centering
\begin{minipage}[h]{4in}
  \begin{algorithmic}[1]
    \STATE \textbf{requires:}\textit{ $\varphi$ is monotone in $\psi$}
    \STATE $\textbf{boolean} \;  \mathtt{isMonVacuous}(\textbf{Formula}\;\varphi,\textbf{Formula}\;\psi,\textbf{Model}\;K)$
    \begin{ALC@g}
      \STATE \textbf{return } $K \models \varphi[\psi\subst \true] \liff 
                                  K \models \varphi[\psi\subst \false]$
    \end{ALC@g}
  \end{algorithmic}
\end{minipage}
  \caption{Vacuity detection algorithm for monotone formulas.}
  \label{fig:monotone-vacuity-algorithm}
\end{figure}

The algorithm for detecting vacuity with respect to monotone
subformulas, called $\mathtt{isMonVacuous}$, is given in
Figure~\ref{fig:monotone-vacuity-algorithm}.  Detecting vacuity of
$\varphi$ with respect to a monotone subformula $\psi$ can be reduced
to comparing the results of two model-checking problems: the one in
which $\psi$ is replaced with \true, and another in which $\psi$ is
replaced with \false. The algorithm is based on the following
intuition. For a fixed model $K$, $\varphi[\psi]$ can be seen as a
monotone function from temporal logic to $\{\true, \false\}$ defined
as follows: $\lambda x \cdot K \models \varphi[\psi \subst x]$. The
formula $\varphi$ is vacuous in $\psi$ if the above function is a
constant (i.e., always \true or always \false). Since the
function is monotone, it is a constant if and only if it assigns the
same value to the extreme points: \true and \false. The correctness of
the algorithm is established by the following theorem.

\begin{theorem}
  \label{thm:monotone-vacuity-algorithm-correct}
  Let $\varphi$ be a temporal logic formula monotone in a subformula
  $\psi$, and $K$ be a \kripke structure. Then
  $\mathtt{isMonVacuous}(\varphi, \psi, K)$ returns \true if and only
  if $\varphi$ is bisimulation vacuous in $\psi$.
\end{theorem}
\begin{proof}
  We first establish the ($\lpmi$) direction. Assume $\varphi$ is
  bisimulation vacuous in $\psi$, and without loss of generality,
  assume that $\varphi$ is satisfied by $K$. From
  Definition~\ref{def:vacuity}, it follows that $\varphi$ holds under
  any interpretation of $\psi$, i.e., $K \models_b \A{x}\varphi[\psi
  \subst x]$. Finally, by specialization, $K \models \varphi[\psi \subst
  \true] \land  K \models \varphi[\psi\subst \false]$.

  For the $(\limp)$ direction, we use the fact that bisimilar
  structures satisfy the same temporal properties. Formally, for a
  formula $\varphi$ with a subformula $\psi$ and a \kripke structure
  $K$,
  \begin{equation*}
    \A{K' \in \Bisim(K)}\A{c \in
      \B} (K \models \varphi[\psi \subst c]) \liff (K' \models \varphi[\psi \subst c])
    \tag*{(constant subst)}
  \end{equation*}
  Furthermore, without loss of generality, we assume that $\varphi$ is
  satisfied by $K$, i.e., 
\begin{equation*}
(K \models \varphi[\psi \subst \true]) \land (K \models \varphi[\psi\subst \false])\,.
\end{equation*} 
The proof proceeds as
  follows:
\[
  \begin{array}{llH}
    & (K \models \varphi[\psi \subst \true]) \land (K \models \varphi[\psi\subst \false])
    & by constant subst\\
\limp   & \A{K' \in \Bisim(K)} (K' \models \varphi[\psi \subst \true]) \land (K' \models \varphi[\psi\subst \false]) & by monotonicity\\
=    & \A{K' \in \Bisim(K)} 
             \A{Y \subseteq S'} K' \models \varphi[\psi \subst Y] & 
           by Definition~\ref{def:qctl-bisimulation}\\
=       & K \models_b \A{x}\varphi[\psi \subst x]
  \end{array}
\]
Hence, by the discussion following Definition~\ref{def:vacuity}, $\varphi$
is bisimulation vacuous in $\psi$.
\end{proof}

From the algorithm $\mathtt{isMonVacuous}$ and
the proof of its correctness, we see that the complexity
of detecting vacuity of monotone formulas is the same as that
for model-checking:

\begin{corollary}
  \label{thm:monotone-vacuity}
  Deciding whether a temporal logic formula $\varphi$ is vacuous in a
  monotone subformula $\psi$ is the same complexity as that of
  model-checking $\varphi$.
\end{corollary}

Note that by itself, the algorithm $\mathtt{isMonVacuous}$
is incomplete since it
requires a user to identify monotonicity of a subformula. However, in
combination with a technique to decide whether a subformula is
monotone, the algorithm leads to a practical and efficient vacuity
detection technique. 

There are several simple syntactic checks to identify monotone
subformulas. For example, if $\psi$ has only a single occurrence in
$\varphi$, then $\varphi$ is monotone in $\psi$, e.g., $\AG(p \lor q
\lor r)$ is monotone in $q$.  Similarly, if $\psi$ is pure in
$\varphi$ (i.e., all occurrences are either positive, like $p$ above,
or negative), then $\varphi$ is monotone in $\psi$.

These simple syntactic checks have already been used in the early work
on vacuity detection by Beer et al.~\cite{beer01} and by Kupferman and
Vardi~\cite{kupferman03}. The algorithms presented in these papers are
equivalent to the algorithm $\mathtt{isMonVacuous}$,
but only apply to
formulas whose monotonicity can be detected syntactically.  
We thus conclude the following:
\begin{theorem}\label{thm:coincide}
All three types of vacuity -- 
syntactic, structure, and bisimulation -- coincide
 for monotone formulas.
\end{theorem}
In particular,  formulas with a single occurrence
of a subformula of interest, or formulas with pure polarity are
(syntactically) monotone.  Thus, by Theorem~\ref{thm:coincide}, the
three definitions of vacuity coincide for such formulas and so
do the algorithms $\mathtt{isMonVacuous}$ and those reported
in \cite{beer01} and \cite{kupferman03}.

It is also interesting to see whether the scope of these simple
syntactic checks for monotonicity can be significantly extended.  We
show that this is not possible in general due to the {\it EXPTIME}-hardness of
this problem.
\begin{theorem}
  \label{thm:deciding-monotonicity}
  Deciding whether a formula $\varphi$ is monotone in a subformula
  $\psi$ is EXPTIME-hard for \ctl, and 2EXPTIME-hard for \ctlast.
\end{theorem}
\begin{proof}
  We reduce the validity problem for \ctl, known to be
  EXPTIME-hard~\cite{fischer79}, to deciding monotonicity. Let
  $\varphi$ be an arbitrary \ctl formula, and $p$ be an atomic
  proposition not occurring in $\varphi$. Then the formula $\psi = (p
  \limp AX p) \lor \varphi$ is monotone in $p$ iff $\varphi$ is
  valid. In general, $\psi$ is not monotone in $p$. However, if
  $\varphi$ is valid, then $\psi$ is valid as well; hence, it is monotone in
  all of its atomic propositions.

  The proof for \ctlast is identical. Note that the validity problem for
  \ctlast is known to be 2EXPTIME-hard~\cite{vardi85}.
\end{proof}

\noindent
Thus, identifying whether a given formula is monotone is as difficult
as vacuity detection in general. It is unlikely that the applicability
of the algorithm $\mathtt{isMonVacuous}$ can be generalized past
syntactically monotone formulas.

In this section, we have studied vacuity detection for monotone
formulas and gave an efficient algorithm for it.  For such
formulas, bisimulation vacuity coincides with syntactic
vacuity.  
  While our algorithm applies to arbitrary monotone
formulas, we have shown that determining whether a given property is
monotone is as hard as the general vacuity detection.
However, for syntactically monotone formulas, such as those with a single
occurrence of a subformula of interest, or formulas with pure polarity,
our algorithm becomes identical to \cite{beer01,kupferman03}.

\subsection{Deciding Vacuous Satisfaction of  \actlast Formulas}

In this section, we present an algorithm for detecting whether an
\actlast formula is satisfied vacuously.   Specifically, 
given an
\actlast formula $\varphi$, a \kripke structure $K$ which is known
to satisfy $\varphi$, and a subformula $\psi$ of $\varphi$, our goal
is to decide whether $\varphi$ is bisimulation vacuous in
$\psi$. We show that this problem is
in the same complexity class as model-checking. This is significant in
practice since properties are often expressed in \actlast or in its
linear fragment, \ltl. By duality, the results of this section extend
to deciding vacuous falsification of \ectlast formulas.

Recall that deciding whether $\varphi$ is satisfied vacuously is
equivalent to model-checking $\A{x} \varphi[\psi \subst x]$ in $K$
under bisimulation semantics. This, in turn, is equivalent to checking
that $\varphi[\psi \subst x]$ is satisfied in every model that is
$x$-bisimilar to $K$.

\begin{figure}[t]
  \centering \input{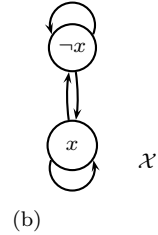}
  \caption{(a) An algorithm for detecting vacuous satisfaction of \actlast formulas, and (b) \kripke structure $\mX$ used by the algorithm.}
  \label{fig:alg-actlast-vacuity}
\end{figure}

Our algorithm for detecting vacuous satisfaction of \actlast formulas
is based on the idea that for \actlast formulas, vacuity detection
can be reduced to a single model-checking instance. The algorithm,
called $\mathtt{isSATVacuous}$,
is shown in Figure~\ref{fig:alg-actlast-vacuity}(a). In the rest
of this section, we first illustrate the algorithm on an example, and
then formally establish its correctness and complexity.

\begin{figure}[t]
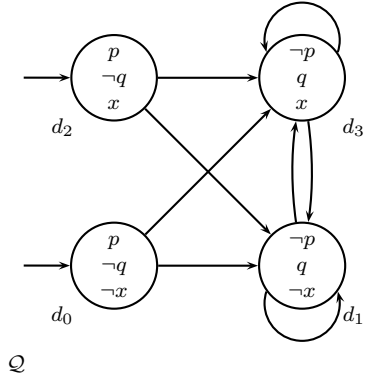

  \centering
  \let\pspicgrid=\pspicture
\begin{pspicgrid}(-1,-1)(4,4)

\rput(-.8,-.8){$\mQ$}
\psset{framesep=0}
\rput(-0.7,0.5){\pnode{init}}
\cnodeput(0.5,0.5){a}{\begin{tabular}{@{}c@{}}$p$\\$ \neg q$\\$\neg x$\end{tabular}}
\cnodeput(3,0.5){b}{\begin{tabular}{@{}c@{}}$\neg p$\\$q$\\$ \neg x$\end{tabular}}

\rput(-0.7,3){\pnode{init2}}
\cnodeput(0.5,3){c}{\begin{tabular}{@{}c@{}}$p$\\$\neg q$\\$ x$\end{tabular}}
\cnodeput(3,3){d}{\begin{tabular}{@{}c@{}}$\neg p$\\$q$\\$ x$\end{tabular}}

\psset{arrows=->}
\ncline{init}{a}
\ncline{a}{b}
\nccircle[angleA=-180]{->}{b}{.5}

\ncline{init2}{c}
\ncline{c}{d}
\nccircle[angleA=0]{->}{d}{.5}

\ncline{a}{d}
\ncline{c}{b}
\ncarc{b}{d}
\ncarc{d}{b}

\nput*{-135}{a}{$d_0$}
\nput*{-45}{b}{$d_1$}
\nput*{-135}{c}{$d_2$}
\nput*{-45}{d}{$d_3$}
\end{pspicgrid}

  \caption{A model $\mQ$ used in the reduction of vacuity detection for  the model $\mP$ from Figure~\ref{fig:samples} to model-checking.} 
  \label{fig:kmin}
\end{figure}

As an example, we consider the problem of detecting whether an
\actlast formula is satisfied vacuously in a  model $\mP$ given in
Figure~\ref{fig:samples}. We show that this problem is reducible to a
single model-checking problem with respect to a model $\mQ$ given in
Figure~\ref{fig:kmin}.  The model $\mQ$ is obtained from $\mP$ by the
following steps: (a) adding a new atomic proposition $x$; (b)
splitting each state of $\mP$ into two states, one interpreting $x$ as
\true and another interpreting $x$ as \false; and (c) adding a
transition between any two states if there is a transition between the
corresponding states of $\mP$.  For example, states $d_0$ and $d_2$ of
$\mQ$ correspond to splitting state $c_0$ of $\mP$; the transition
between $d_2$ and $d_1$ in $\mQ$ corresponds to the transition between
$c_0$ and $c_1$ in $\mP$; and there is no transition between $d_0$ and
$d_2$ in $\mQ$ since there is no corresponding self-loop on $c_0$ in
$\mP$.

It is easy to see that $\mQ$ is $x$-bisimilar to $\mP$: $\mQ$ differs
from $\mP$ only in its interpretation of the new variable $x$, but
otherwise has all of the same behaviors.  Furthermore, $\mQ$ does not
enforce any temporal constraints on $x$ -- from any state, $x$ can
evolve to either \true or \false. Thus, $\mQ$ can simulate (i.e.,
match the behavior of) any \kripke structure that is $x$-bisimilar to
$\mP$. For example, $d_0$ can simulate any state that is $x$-bisimilar
to $c_0$, and $d_2$ can simulate any state that is $x$-bisimilar to
$c_1$.  Recall that simulation preserves satisfaction of \actlast
formulas (Theorem~\ref{thm:simulation-actlast}). Thus, since $\mQ$
simulates every structure that is $x$-bisimilar to $\mP$, it satisfies
an \actlast formula \emph{if and only if} the formula is satisfied by
every structure $x$-bisimilar to $\mP$. This reduces model-checking a
formula $\varphi[\psi\subst x]$ on \emph{all} structures that are
$x$-bisimilar to $\mP$ to a \emph{single} model-checking problem on
$\mQ$!  Hence, checking whether $\varphi$ is $\psi$-vacuous on $\mP$
is equivalent to model-checking $\varphi[\psi\subst x]$ on $\mQ$.

While  $\mQ$ has twice as many states as $\mP$, both
structures share the same symbolic representation of the transition
relation, represented by the formula
\begin{equation*}
  (p \land \neg q \land \neg p' \land q') \lor (\neg p \land
  q \land \neg p' \land q').
\end{equation*}
This means that for a symbolic model-checking algorithm,
checking $\mQ$ and a seemingly smaller model $\mP$ is equally
easy (or equally hard).

We now return to the algorithm $\mathtt{isSATVacuous}$.  This algorithm uses a
\kripke structure $\mX$ shown in
Figure~\ref{fig:alg-actlast-vacuity}(b) and defined
as 
$\mX \eqdef (AP^{\mX}, S^{\mX}, S_0^{\mX}, R^{\mX},
I^{\mX})$, with
a single atomic proposition $x$ ($AP^{\mX} = \{x\}$), two
states ($S^{\mX} = \{0, 1\}$),
all states being initial
($S_0^{\mX} = S^{\mX}$), 
any transition being allowed 
($R^{\mX} = S^{\mX} \times S^{\mX}$), 
and $x$ being interpreted as $I^{\mX}(0, x) = \false$
and $I^{\mX}(1, x) = \true$.

The correctness of the algorithm is based on the observation that for
any \kripke structure $K$, the parallel synchronous composition $K
\mathbin{||} \mX$ of $K$ and $\mX$ (assuming that $x$ is a fresh variable
for $K$) simulates any structure $K'$ that
is $x$-bisimilar to $K$. 

\begin{theorem}
  \label{thm:x-simulation}
  Let $K = (AP, S, R, S_0, I)$ be an arbitrary \kripke structure, and
  $K'=(AP \cup \{x\}, S', R', S'_0, I')$ be $\{x\}$-bisimilar to $K$.
  Then $K'$ is simulated by $K \mathbin{||} \mX$.
\end{theorem}
\begin{proof}
  By Definition~\ref{def:sync-composition}, the \kripke
  structure $K \mathbin{||} \mX$ is \[(AP \cup
  \{x\}, S\times \{0,1\}, S_0 \times \{0,1\}, R^x, I^x),\] where $R^x
  (\langle s, i \rangle, \langle t, j \rangle) \liff R(s,t)$, and
  \[
  I^x(\langle s, i \rangle, p) = \begin{cases}
    I(s, p) & \text{if } p \neq x\\
    I^{\mX}(i, x) & \text{if } p = x.
  \end{cases}
  \]

  Let $\rho \subseteq S \times S'$ be the $\{x\}$-bisimulation
  relation between $K$ and $K'$. We claim that $K \mathbin{||} \mX$
  simulates $K'$ via the relation \[\rho^{x} = \{ ( \langle s, i
  \rangle, t) \mid \rho(s,t) \land I^{x}(\langle s, i \rangle, x) =
  I'(t, x)\}.\]
  
  From the definition of $\rho^{x}$, it follows immediately that $\rho^{x}(
  \langle s, i \rangle, t) \liff (I^{x}(\langle s, i \rangle ) =
  I(t))$, thus, it satisfies the first condition of simulation.  The
  proof of the second condition is given below:
  \[
  \begin{array}{ll}
         & \rho^{x}( \langle s, i\rangle, t) \land R'(t, t') \\
\limp    & \lhint{since $K'$ is $\{x\}$-bisimilar to $K$}\\
         & \E{s' \in S} \rho(s', t') \land R(s, s')\\
\limp    & \lhint{by definition of $K \mathbin{||} \mX$}\\
         & \E{s' \in S} \A{j \in \{0,1\}} R^x(\langle s, i \rangle, \langle s', j \rangle) \land \rho(s', t')\\
\limp    & \lhint{since $I'(t', x) \in \B$}\\
         &  \E{s' \in S} \E{j \in \{0,1\}} R^x(\langle s, i \rangle, \langle s', j \rangle) \land \rho(s', t') \land I^x(\langle s', j\rangle) = I'(t')\\
\limp    & \lhint{by definition of $\rho{^x}$}\\
         & \E{s' \in S} \E{j \in \{0,1\}} R^x(\langle s, i \rangle, \langle s', j \rangle) \land \rho^{x}(\langle s', j \rangle, t') 
  \end{array}
  \]
  Finally, if $t$ is an initial state of $K'$, then there exists an $i
  \in \{0,1\}$ such that $\rho^{x}( \langle s, i\rangle, t)$ holds,
  which establishes that $K \mathbin{||} \mX$ simulates $K'$ via
  $\rho^{x}$.
\end{proof}

Since simulation preserves \actlast, vacuity detection for an
arbitrary \actlast formula is reducible to model-checking over
$K\mathbin{||} \mX$.  This proves correctness of
$\mathtt{isSATVacuous}$.

\begin{proposition}
  \label{thm:alg-actlast-vacuity-correctness}
  Let $\varphi$ be an \actlast formula with a subformula $\psi$, $K$
  be a \kripke structure, and assume that $K$ satisfies $\varphi$.
  Then  $\mathtt{isSATVacuous}(\varphi,\psi,K)$ returns \true if
  and only if $\varphi$ is bisimulation vacuous in $\psi$.
  
\end{proposition}
\begin{proof}
  Let $\varphi$ be a formula satisfied by $K$. We show that $\varphi$
  is $\psi$-vacuous iff the formula \mbox{$\varphi[\psi \subst x]$} is
  satisfied by $K \mathbin{||} \mX$.  

Since $K \mathbin{||} \mX$ is
  $\{x\}$-bisimilar to $K$, the proof of ($\limp$) direction is
  trivial. 

For ($\lpmi$) direction, if $\varphi[\psi \subst x]$ holds
  in $K \mathbin{||} \mX$, then by Theorem~\ref{thm:x-simulation} and
  Theorem~\ref{thm:bisimulation-ctlast} it holds in every
  $\{x\}$-bisimulation of $K$.
\end{proof}

An immediate consequence of
Proposition~\ref{thm:alg-actlast-vacuity-correctness} is that for the LTL
fragment of \actlast, our bisimulation vacuity is equivalent to trace
vacuity of Armoni et al~\cite{armoni03}. That is, for any fixed model
$K$, an LTL formula $\varphi$ is trace vacuous in $\psi$ if and
only if it is bisimulation vacuous in $\psi$. We further elaborate on
this connection between the two definitions in
Section~\ref{sec:related-work}.

From the algorithm $\mathtt{isSATVacuous}$
and the
proof of its correctness, it is easy to see that the complexity of
detecting vacuous satisfaction of \actlast formulas is in the same
complexity class as model-checking:

\begin{corollary}
  \label{thm:actlast-vacuity-complexity}
  Let $\varphi$ be an \actlast formula, $\psi$ be a subformula of
  $\varphi$, and $K$ be a \kripke structure. Deciding whether $K$
  satisfies $\varphi$ $\psi$-vacuously is in the same complexity class
  as model-checking $\varphi$.
\end{corollary}

\noindent
As mentioned earlier, while the explicit statespace of $K \mathbin{||}
\mX$ is twice of that of $K$, $K \mathbin{||} \mX$ does not impose any
restrictions on the atomic proposition $x$; therefore, the symbolic
representation of its transition relation is identical to that of $K$.

In this section, we described an algorithm, $\mathtt{isSATVacuous}$,
for detecting whether an \actlast formula is satisfied vacuously.
We proved correctness of this algorithm and showed that
checking whether an \actlast formula $\varphi$ is
vacuous in some subformula is no more expensive  
than model-checking $\varphi$.

\subsection{Deciding Vacuous Satisfaction of \ctlast in Universal Subformulas}
\label{sec:reduct-ctlast-form}

In the rest of this section, we show that the algorithm
$\mathtt{isSATVacuous}$ can be used not only for detecting vacuous
satisfaction of \actlast formulas but also for detecting vacuous
satisfaction of \ctlast formulas with respect to universal
subformulas.  That is, under the assumption that a subformula $\psi$
occurs only under universal path quantifiers in the negation normal
form of $\varphi$, $\mathtt{isSATVacuous} (\varphi, \psi, K)$ returns
\true if and only if $\varphi$ is satisfied vacuously in $\psi$.

  Given a fixed model $K$, the structure of a temporal formula
  $\varphi$ can be simplified by replacing state subformulas with
  propositional expressions without affecting the satisfiability of
  $\varphi$. For example, consider the model $\mO$ in
  Figure~\ref{fig:samples}(b) and the property $AF EG p$. For this
  model, formula $EG p$ can be simplified to $p$, and formula
  \mbox{$AF EG p$ --- to $AF p$.} We use the notation $Prop(\varphi,
  K)$ to denote \emph{some} such propositional simplification of
  $\varphi$ with respect to a model $K$.  Formally, $Prop(\varphi, K)$
  is a formula obtained by replacing some state subformula $\psi$ of
  $\varphi$ with a propositional encoding of the set $||\psi||^K$ of
  all the states of $K$ that satisfy $\psi$.

Propositional simplification does not affect satisfaction. That is, a
structure $K$ satisfies $\varphi$ if and only if it satisfies a
propositional simplification of $\varphi$:
\begin{equation*}
  K \models \varphi \liff K \models Prop(\varphi,K) 
                \tag*{(propositional simplification)}
\end{equation*}
Moreover, this property is preserved by bisimulation -- if $K$ and
$K'$ are bisimilar, then $\varphi$ can be simplified with respect to
either model without affecting its satisfaction on both models. That
is, $K'$ satisfies $\varphi$ if and only if it satisfies any
propositional simplification of $\varphi$ with respect to a bisimilar
model $K$.

\begin{theorem}
  \label{thm:bisim-prop-subst}
  Let $K$ and $K'$ be two structures such that $K$ is $x$-bisimilar to
  $K'$ via a relation $\rho$, and let $\varphi$ be a \ctlast formula
  not containing $x$. Then $K'$ satisfies $\varphi$ iff it satisfies a
  propositional simplification of $\varphi$ with respect to $K$:
  \begin{equation*}
    K' \models \varphi \liff K' \models Prop(\varphi,K)\,.
  \end{equation*}
\end{theorem}
\begin{proof}
  Let $S$ and $S'$ denote the statespaces and $s_0$ and $s'_0$ denote
  the initial states of $K$ and $K'$, respectively. Then
  \[
  \begin{array}{llH}
    & K' \models \varphi & definition of $\models$\\ 
=   & K',s'_0 \models \varphi  & property of $\rho$\\
=   & \E{s\in S}  K,s \models \varphi \land (s,s'_0) \in \rho 
                                         & propositional simplification\\
=   & \E{s \in S} K,s \models Prop(\varphi,K) \land (s,s'_0) \in \rho 
                                      & property of $\rho$\\
=   & K',s'_0 \models Prop(\varphi,K) & definition of $\models$\\
=   & K' \models Prop(\varphi,K)\\
  \end{array}
  \]
\end{proof}

We now use Theorem~\ref{thm:bisim-prop-subst} to establish the main
theorem of this section. We show that for any fixed
model, a \ctlast formula with a universal subformula $\psi$ can be
turned into an \actlast formula without affecting the $\psi$-vacuity
of the formula.

\begin{theorem}
  \label{thm:alg-ctlast-vacuity-correctness} 
  Let $\varphi$ be a \ctlast formula with a universal subformula
  $\psi$, and $K$ be a \kripke structure. Assume that $K$
  satisfies $\varphi$. Then $\mathtt{isSATVacuous}(\varphi, \psi, K)$
  returns \true if and only if $\varphi$ is bisimulation vacuous in
  $\psi$.
\end{theorem}
\begin{proof}
  Let $\mX$ be a \kripke structure as depicted in
  Figure~\ref{fig:alg-actlast-vacuity}(b). We show that for a \kripke
  structure $K$, $\varphi[\psi \subst x]$ is satisfied by $K || \mX$
  iff $\varphi$ is vacuous in $\psi$. 

The ``if'' direction is trivial.

For the ``only if'' direction, assume that $K || \mX$ satisfies
$\varphi[\psi \subst x]$. Let $Prop(\varphi[\psi \subst x],K)$ be the
result of replacing all existential state subformulas of $\varphi[\psi
\subst x]$ with their propositional simplification in $K$.  Since
$\psi$ occurs only in the scope of universal quantifiers, these
subformulas do not contain $x$ and can be interpreted on $K$. By
Theorem~\ref{thm:bisim-prop-subst}, $K || \mX$ satisfies
$Prop(\varphi[\psi \subst x],K)$.  Since $\psi$ is universal,
$Prop(\varphi[\psi \subst x], K)$ is in \actlast.  Applying
Theorem~\ref{thm:x-simulation} and then
Theorem~\ref{thm:simulation-actlast}, we get that every \kripke
structure $K'$ that is $x$-bisimilar to $K$ satisfies
$Prop(\varphi[\psi \subst x], K)$. By
Theorem~\ref{thm:bisim-prop-subst}, $K'$ satisfies $\varphi[\psi
\subst x]$ as well. Hence, by Definition~\ref{def:vacuity}, $\varphi$
is bisimulation vacuous in $\psi$.
\end{proof}

Theorem~\ref{thm:alg-ctlast-vacuity-correctness} implies that 
detecting whether an arbitrary \ctlast formula is
satisfied vacuously in a universal subformula is in the same
complexity class as model-checking:
\begin{corollary}
  \label{thm:alg-ctlast-vacuity-complexity}
  Let $\varphi$ be a \ctlast formula, $\psi$ be a universal
  subformula of $\varphi$, and $K$ be a \kripke structure. Deciding
  whether $K$ satisfies $\varphi$ $\psi$-vacuously is in the same
  complexity class as model-checking $\varphi$.
\end{corollary}

In this section, we have shown that the algorithm
$\mathtt{isSATVacuous}$ is applicable not only to \actlast formulas, but also
to detecting vacuous satisfaction of arbitrary \ctlast formulas in
universal subformulas.  We have also shown that 
vacuity detection for this more general case remains in the same complexity class as
model-checking.

\section{Vacuity and Abstraction}
\label{sec:vacuity-abstraction}
The statespace explosion problem, i.e., the fact that
the size of a model doubles with an addition of each
new atomic proposition, is one of the major challenges in practical
applications of model checking. Abstraction is the most popular and
most effective technique to combat this problem.  In this section, we
explore the interactions between vacuity detection and abstraction. 

\subsection{Abstraction}
\label{sec:abstraction}
The key principle of abstraction is to replace model checking of a given
property $\varphi$ on a \emph{concrete} model $K_c$ with model
checking of the property on an \emph{abstract} model $K_\alpha$. The
abstract model $K_\alpha$ is typically chosen such that it is smaller
and/or easier to represent symbolically than $K_c$.

Here, we consider the two most commonly used abstractions. In a
\emph{bisimulation-based abstraction}, the abstract model $K_\alpha$
is required to be bisimilar to the concrete model $K_c$. Cone of
influence~\cite{clarke99} and symmetry
reduction~\cite{clarke98,owei05} are two prominent examples of
bisimulation-based abstraction. This abstraction is \emph{sound and
  complete} for \ctlast. That is, if a given property is satisfied or
refuted by the abstract model, then it is, respectively, satisfied or
refuted by the concrete one as well.

In a \emph{simulation-based abstraction}, the abstract model
$K_\alpha$ is required to simulate the concrete model $K_c$. This is
the most commonly used abstraction technique for hardware and
software model checking, e.g.,~\cite{graf97,ball01}. Simulation-based
abstraction is \emph{sound (but incomplete)} for \actlast. That is,
the abstract model over-approximates the behaviors of the concrete
one. Thus, if an \actlast property is satisfied by $K_\alpha$, it is
satisfied by $K_c$, but the converse is not true in general.

\subsection{Vacuity Detection in the Presence of Abstraction}
\label{sec:vac-det-in-abs}
In this section, we explore the preservation of vacuity for
bisimulation- and simulation-based abstractions. Clearly,
bisimulation-based abstraction is sound and complete for vacuity of
\ctlast.

\begin{proposition}
  Let $K_\alpha$ and $K_c$ be \kripke structures such that $K_\alpha$
  is a bisimulation-based abstraction of $K_c$, and let $\varphi$ be a
  \ctlast formula with a subformula $\psi$. Then, $\varphi$ is
  $\psi$-vacuous in $K_a$ iff $\psi$-vacuous in $K_c$.
\end{proposition}
\begin{proof}
  By definition of bisimulation-based abstraction, $K_\alpha$ and
  $K_c$ are bisimilar. By
  Proposition~\ref{thm:bisimilarity-and-vacuity}, bisimulation vacuity
  is invariant under bisimulation.
\end{proof}

Note that bisimulation-based abstraction is \emph{not} sound with
respect to alternative definitions of vacuity! An example in
Section~\ref{sec:vacuity-definition} (Weakness 2) shows that the
abstraction is not sound with respect to syntactic vacuity: the model
$\mO$ can be viewed as an abstraction of a concrete model $\mL$. Then,
property $P_4$ is vacuous in the concrete model, but is
non-vacuous in the abstract, i.e., abstraction has masked vacuity.
An
example in Section~\ref{sec:structure-vacuity} illustrates a similar
situation for structure vacuity: the model $\mL$ can be viewed as
concrete and the model $\mM$ as abstract. Property $P_4$ 
is vacuous in the concrete model and non-vacuous in the
abstract.

We now turn our attention to simulation-based abstraction.
Recall that this abstraction is only sound for
\actlast and thus we can only expect it to be sound for vacuity of
\actlast properties. Furthermore, this abstraction is not complete and
thus
we do not expect it to be complete for vacuity either.  We show that
below.

\begin{theorem}
  \label{thm:simulation-and-vacuity}
  Let $K_\alpha$ and $K_c$ be two \kripke structures such that
  $K_\alpha$ simulates $K_c$, and let $\varphi$ be an \actlast formula
  with a subformula $\psi$. Then, whenever $\varphi$ is $\psi$-vacuous
  in $K_a$, it is $\psi$-vacuous in $K_c$.
\end{theorem}
\begin{proof}
  By Proposition~\ref{thm:alg-actlast-vacuity-correctness}, $\varphi$
  is $\psi$-vacuous in $K_\alpha$ iff $(K_\alpha \mathop{||} \Chi)
  \models \varphi[\psi \subst x]$, where $\Chi$ is the \kripke
  structure shown in Figure~\ref{fig:alg-actlast-vacuity}(b). 
  $K_\alpha$ simulates $K_c$ and thus $K_\alpha \mathop{||} \Chi$
  simulates $K_c \mathop{||} \Chi$ as well. Since simulation preserves
  \actlast, $K_c \mathop{||} \Chi \models \varphi[\psi \subst x]$. By
  Proposition~\ref{thm:alg-actlast-vacuity-correctness}, $\varphi$ is
  $\psi$-vacuous in $K_c$.
\end{proof}

Soundness of simulation-based abstraction with respect to vacuity of
\actlast is a trivial corollary of
Theorem~\ref{thm:simulation-and-vacuity}:
\begin{corollary}
  Let $K_\alpha$ and $K_c$ be \kripke structures such that $K_\alpha$
  is a simulation-based abstraction of $K_c$, and let $\varphi$ be a
  \actlast formula with a subformula $\psi$. Then, whenever $\varphi$
  is $\psi$-vacuous in $K_a$, it is $\psi$-vacuous in $K_c$.
\end{corollary}

The converse of Theorem~\ref{thm:simulation-and-vacuity}
is not true. As a counterexample, consider two structures, $\mV$ and
$\mV_\alpha$, shown in Figure~\ref{fig:alg-actlast-vacuity} and a
property
\[
P_8 \eqdef \AG \left( p \limp \AX q \right)\,.
\]
While $P_8$ is satisfied vacuously in $\mV$, it is non-vacuous in
$\mV_\alpha$. Thus, vacuity of a formula might be ``hidden'' by
abstraction.

\begin{figure}[t]
  \centering
  \let\pspicgrid=\pspicture
\begin{pspicgrid}(0,0)(12,3)
\rput[bl](0,0){%
\begin{pspicture}(-1,0)(4,3)

\rput[bl](-.8,0){$\mV$}
\psset{framesep=0}
\rput(-0.3,0.5){\pnode{init}}
\cnodeput(0.5,0.5){a}{\begin{tabular}{@{}c@{}}$\neg p$\\$ q$\end{tabular}}
\cnodeput(3,0.5){b}{\begin{tabular}{@{}c@{}}$\neg p$\\$\neg q$\end{tabular}}

\cnodeput(0.5,2){c}{\begin{tabular}{@{}c@{}}$p$\\$q$\end{tabular}}
\cnodeput(3,2){d}{\begin{tabular}{@{}c@{}}$p$\\$\neg q$\end{tabular}}

\psset{arrows=->}
\ncline{init}{a}
\ncline{a}{b}
\nccircle[angleA=-180]{->}{b}{.3}

\end{pspicture}}

\rput[bl](6,0){%
\begin{pspicture}(-1,0)(4,3)

\rput[bl](-.8,0){$\mV_\alpha$}
\psset{framesep=0}
\rput(-0.3,0.5){\pnode{init}}
\cnodeput(0.5,0.5){a}{\begin{tabular}{@{}c@{}}$\neg p$\\$ q$\end{tabular}}
\cnodeput(3,0.5){b}{\begin{tabular}{@{}c@{}}$\neg p$\\$\neg q$\end{tabular}}

\cnodeput(0.5,2){c}{\begin{tabular}{@{}c@{}}$p$\\$q$\end{tabular}}
\cnodeput(3,2){d}{\begin{tabular}{@{}c@{}}$p$\\$\neg q$\end{tabular}}

\psset{arrows=->}
\ncline{init}{a}
\ncline{a}{b}
\nccircle[angleA=-180]{->}{b}{.3}

\ncline{c}{d}
\nccircle[angleA=0]{->}{d}{.3}

\ncline{a}{d}
\ncline{c}{b}
\ncarc{b}{d}
\ncarc{d}{b}

\end{pspicture}}

\end{pspicgrid}
  \caption{A concrete \kripke structure $\mV$ and its existential
    abstraction $\mV_\alpha$.}
  \label{fig:vacabs}
\end{figure}
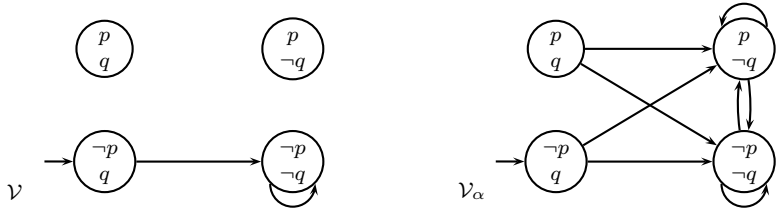

Note that simulation-based vacuity is \emph{not} sound with
respect to syntactic and structural definitions of vacuity. Same
examples as used for bisimulation vacuity above apply here as well
since the property $P_4$ is in \actlast.

In summary, we showed that bisimulation vacuity interacts well with
two most common abstraction techniques. Bisimulation-based abstraction
is sound and complete for \ctlast and is also sound and complete for
vacuity. On the other hand, simulation-based abstraction is sound (but
incomplete) for \actlast and is only sound (but incomplete) for
bisimulation vacuity. Moreover, neither of the abstractions is sound with respect
to syntactic or structure vacuity.

Combining vacuity and abstractions other than simulation-based and
bisimulation-based (such as the mixed-simulation-based abstraction of Dams et al.~\cite{dams97} which we studied in conjunction with vacuity in \cite{gurfinkel04a})
would require similar reasoning as described in this section but is beyond
the scope of this paper. 

\section{Related Work}
\label{sec:related-work}

In this section, we survey related work. We begin by a general
overview of vacuity research that is based on the (modifications of)
the original syntactic vacuity of Beer et al.~\cite{beer97}. We then
give an in-depth comparison between bisimulation vacuity and trace
vacuity of Armoni et al.~\cite{armoni03}. We conclude this section by
a discussion of other sanity checks to complement model-checking.

\heading{Syntactic Vacuity.} The majority of the work on vacuity is
based on the definition of syntactic vacuity (see
Definition~\ref{def:formula-vacuity}) of Beer et al.~\cite{beer97}.
This definition and the corresponding vacuity detection algorithm have
been extended and adapted to a variety of property languages: to \ctlast
in~\cite{kupferman99}, to the modal $\mu$-calculus in~\cite{dong02},
to temporal logic with regular expressions in~\cite{bustan05}, and to
the logic of symbolic trajectory evaluation in~\cite{tzoerf06}.

The notion of syntactic vacuity has been extended in a variety of
ways.  Gurfinkel and Chechik~\cite{gurfinkel04a} and Chockler and
Strichman~\cite{chockler07,chockler09} have studied \emph{mutual
  vacuity} that considers vacuity in several subformulas
simultaneously. Dong et al.~\cite{dong02} and independently Samer and
Veith~\cite{samer04} have explored a notion of vacuity in which a
weaker formula (such as $AF p$) can be replaced by a stronger one
(such as $AX p$). Chockler and Strichman~\cite{chockler07,chockler09}
have also explored vacuity between multiple properties, independently
of a model.

Several modifications to the na\"ive vacuity detection algorithms of
Beer et al.~\cite{beer97} and Kupferman and Vardi~\cite{kupferman99}
have been proposed. Purandare and Somenzi~\cite{purandare02} use the
parse tree of temporal formula to enable information sharing between
vacuity detecting passes of a symbolic model-checker.  Gurfinkel and
Chechik~\cite{gurfinkel04a} give an algorithm, based on multi-valued
model-checking, that detects all instances of vacuity of a formula in
a single pass. Simmonds et al.~\cite{simmonds09} use resolution proofs
to speed up vacuity detection for bounded SAT-based model-checking.

\heading{Semantic Vacuity.}  We were inspired by the work of Armoni et
al.~\cite{armoni03}.  In \cite{armoni03}, the authors show many
anomalies of the syntactic approach to vacuity, and informally argue
for a set of robustness criteria. They present a semantic definition
of vacuity for LTL, called \emph{trace vacuity}, and develop an
algorithm for detecting it. In this article, we build on this work by
formalizing the criteria for robust vacuity using bisimulation, and by
extending semantic vacuity to branching-time logic.

In what follows, we give an in-depth comparison between bisimulation
vacuity that is introduced in this article and trace vacuity of Armoni
et al. We give a formal definition of trace vacuity and its trivial
extension to \ctlast and show that this extension is not robust. We
then show that bisimulation vacuity is a proper extension of trace
vacuity by proving that they coincide for LTL properties. Finally, we
compare the algorithms for detecting trace vacuity for LTL and
bisimulation vacuity for \actlast.

Originally, trace vacuity was defined using tree semantics of QTL,
making it directly applicable to \ctlast. A formal definition is given below:
\begin{definition}\cite{armoni03}
  \label{def:trace-vacuity}
  A temporal logic formula $\varphi$ is trace $\psi$-vacuous in a
  \kripke structure $K$ if and only if $K \models_T \A{x} \varphi[\psi \subst
  x]$.
\end{definition}

However, the following example illustrates that trace vacuity is not
robust for branching temporal logic.  Consider the property $(AX p
\lor AX \neg p)$.  It is trace $p$-vacuous in the model $\mL$ in
Figure~\ref{fig:k-one} and not trace $p$-vacuous in the model $\mM$ in
Figure~\ref{fig:xvariants}.  Recall that these two models are
bisimilar and thus should behave identically with respect to
vacuity.  Thus, trace vacuity is not robust:  when applied to branching
time properties, it becomes sensitive to irrelevant changes to the
model (i.e., it suffers from ``Weakness 2'' as described in
Section~\ref{sec:vacuity-definition}). %

Bisimulation vacuity is a proper extension of trace vacuity: i.e., trace
and bisimulation vacuity coincide for LTL. Formally, if an LTL formula
$\varphi$ is trace vacuous with respect to a structure $K$, then
$\varphi$ is bisimulation vacuous w.r.t. $K$ as well, and vice versa.
\begin{theorem} 
  Let $\psi$ be a path formula (i.e., expressed in LTL), and $x$ be an atomic
  proposition occurring in $\psi$. Then, tree and bisimulation
  semantics of quantified temporal formula $\A{x} A \psi$ are
  equivalent. Formally, for any model $K$,
  \begin{equation*}
    K \models_T \A{x} A\psi \liff K \models_b \A{x} A\psi\,.
  \end{equation*}
\end{theorem}
\begin{proof}
  The ``$\lpmi$'' direction follows from
  Theorem~\ref{thm:qctl-sem-reln}. We prove the ``$\limp$'' direction by
  contradiction. Assume that $K \models_T \A{x} A\psi$, and
  $K \not\models_b \A{x} A \psi$. By the assumption, any trace that
  is an $x$-variant of a trace of $K$ satisfies the path formula
  $\psi$. Furthermore, there exists a structure $K'$ $x$-bisimilar to
  $K$ with a trace $\pi \in K'$ such that $\pi$ violates the path
  formula $\psi$, i.e., $\pi \not\models \psi$. However, $\pi$ also
  belongs to some $x$-variant of a tree unrolling $T(K)$ of $K$.
  Hence, $\pi \models \psi$, which contradicts the assumption.
\end{proof}

$\mathtt{isSATVacuous}$, the algorithm  for detecting vacuous
satisfaction of \actlast formulas presented in this article, is very
similar to the one suggested by Armoni et al. for detecting trace
vacuity for \ltl. The main difference is that our algorithm is based
on changing the model and does not impose any restrictions on the
model-checking procedure to be used. In contrast, the algorithm of
Armoni et al.  is based on changing the automaton corresponding to the
\ltl formula and depends on an automata-theoretic model-checking
procedure.  Both of the algorithms can be used interchangeably for
\ltl formulas and have the same time and space complexity.

\heading{Proof-based vacuity.} In~\cite{namjoshi04}, Namjoshi has
introduced a proof-based variant of vacuity. Although it is called
\emph{proof vacuity} in the original paper, we  refer to it as
\emph{forall-proof vacuity}. The key idea behind this vacuity
is to examine the proofs of $K \models \varphi$ for a \kripke
structure $K$ and a formula $\varphi$. Informally, a formula $\varphi$
is forall-proof vacuous in a subformula $\psi$ if $\psi$ is not used
in any proof of $K \models \varphi$. Of course, a formal definition
depends on the exact interpretation of the notion of ``proof''. In comparison, other
definitions of vacuity, as well as bisimulation vacuity considered
here, are of the ``existential'' nature: a formula is vacuous if there
exists a ``proof'' that does not use a subformula.

The forall-proof vacuity is semantic. We conjecture that it is
invariant under bisimulation since model-checking proofs can be lifted
through a bisimulation relation. This would make the forall-proof vacuity
robust in the sense of this article, and more strict compared with
bisimulation vacuity. We also conjecture that in this case,
exists-proof vacuity may coincide with bisimulation vacuity. At the
moment, both of these conjectures remain open.

Exist-proof vacuity has been explored in the context of SAT-based
bounded model checking (BMC)~\cite{simmonds09}. One of the interesting
results of this paper is that it is possible that a formula $\varphi$
be bisimulation vacuous in $\psi$ in a model $K$, yet there is no
\emph{resolution proof} of bounded satisfaction of $K \models \varphi$
that does not use $\psi$. This follows from the fact that resolution
proofs are syntactic (while the proofs used in~\cite{namjoshi04} are
semantic) and may include ``semantically-useless'' resolutions.

\heading{Beyond vacuity.} Vacuity detection can be seen as a ``sanity
check''. It provides the user with an additional degree of confidence
that the result of the model-checking is not trivial. Another useful
sanity check is \emph{coverage}: detecting which part of the model was
responsible for property satisfaction. It was shown by
Kupferman~\cite{kupferman06} that the two problems are closely related
and that techniques for one problem can be adapted for the other.

Perhaps more surprisingly, vacuity detection is also closely connected
to 3-valued abstraction~\cite{gurfinkel05b}.  The two techniques have
dual goals: in vacuity, we check whether any part of the formula can
be simplified or ``abstracted away''; in abstraction, we look for
parts of the model that can be removed without affecting satisfaction
of its properties.  In particular, in~\cite{gurfinkel05b}, we use the
theoretical developments from this article to identify when
thorough~\cite{bruns00} and compositional semantics of 3-valued
model-checking coincide.

In this article, we have considered vacuity only from the perspective
of the property expressed in temporal logic. A more refined vacuity,
or a sanity check, is possible when additional information about the
intended meaning of a property is available. For example, Chechik et
al.~\cite{chechik07} use an assumption that the verification problem
includes a combination of a system and an environment. With this
assumption, they present a sanity check that detects whether a formula
is established solely by the environment. Ben-David et
al.~\cite{bendavid07} assume that a property has a well defined pre-
and post-condition, and present a more refined vacuity check aimed to
find formulas whose pre-conditions are never satisfied. %

\section{Conclusion}
\label{sec:conclusion}
Dealing with vacuous or meaningless satisfaction of properties is a
recognized problem in practical applications of automated
verification.  Over the years, a number of researchers have attempted
to formally capture this notion, calling it \emph{vacuity}.  In this
article, we presented \emph{bisimulation vacuity} as a uniform
definition of vacuity for both branching and linear temporal logics.
Bisimulation vacuity extends syntactic vacuity of Beer et
al.~\cite{beer97} to subformulas of mixed polarity, and extends trace
vacuity of Armoni et al.~\cite{armoni03} to branching temporal logics.
Following Armoni et al.~\cite{armoni03}, we showed that bisimulation
vacuity is \emph{robust}, i.e., independent of logic embedding and
of trivial changes to the model, and enjoys all of the
advantages of trace vacuity.  We also showed that for many important
fragments of temporal logic, vacuity detection is reducible to
model-checking, and thus leads to simple and practical
implementations. In particular, this applies to deciding whether a
\ctlast formula is satisfied vacuously in a universal subformula.
We then explored the preservation of vacuity by abstraction.

The contributions of our work are two-fold. From the theoretical
perspective, we studied the complexity of vacuity detection and showed
that for branching-time logics, it is as hard  as the satisfiability
problem. That is, vacuity detection is exponentially more expensive
than model-checking. This implies that in general vacuity detection
is not computationally tractable, and there does not exist a simple,
practical vacuity detection algorithm for the entire logic.

From the practical perspective, we have identified fragments of
temporal logics for which vacuity can be detected effectively, and
provided the corresponding vacuity detection algorithms.
Specifically, for these fragments, our algorithms are very similar to
the one studied by Armoni et al.~\cite{armoni03}. Thus, we know that
they are effective in practice for checking vacuity of LTL properties.
Since the publication of the conference version of this paper,
\cite{gurfinkel04b}, we have done further studies with our definition
of bisimulation vacuity, implementing it in the setting of bounded
model-checking~\cite{simmonds09} and applying it to the IBM Formal
Verification Benchmarks Library~\cite{cnf07}.

\begin{acks}
  A preliminary version of many of the ideas discussed in this work
  has appeared in~\cite{gurfinkel04b}.  We are grateful to anonymous
  referees of FMCAD'04 for helping improve the presentation and
  technical clarity of this paper.  We also thank K. Namjoshi for
  insightful discussions. This work was supported in part by NSERC,
  OGS, and IBM.
\end{acks}

{{\bibliographystyle{acmtrans}
\bibliography{combined}}}

\begin{received}
Received Month Year; revised Month Year; accepted Month Year
\end{received}
\end{document}